\documentclass[11pt,a4paper]{article}

\usepackage{color}

\font\tenrsfs=rsfs10 at 12pt
\font\sevenrsfs=rsfs7
\font\fiversfs=rsfs5
\newfam\rsfsfam

\textfont\rsfsfam=\tenrsfs
\scriptfont\rsfsfam=\sevenrsfs
\scriptscriptfont\rsfsfam=\fiversfs
\def\mathscr#1{{\fam\rsfsfam\relax#1}}

\def\circa#1{\,\raise.3ex\hbox{$#1$\kern-.75em\lower1ex\hbox{$\sim$}}\,}

\newcommand{\myR}{{\rm I\!R}}

\usepackage{amsmath}
\usepackage{amsfonts}
\usepackage{amssymb}

\usepackage[hyperfootnotes=false,hidelinks]{hyperref}
\usepackage{graphicx}

\newcommand{\eq}[1]{(\ref{#1})}
\newcommand{\be}{\begin{equation}}
\newcommand{\ee}{\end{equation}}
\newcommand{\bea}{\begin{eqnarray}}
\newcommand{\ena}{\end{eqnarray}}
\newcommand{\no}{\noindent}

\newcommand\m{\ensuremath{\mu}}

\newcommand\n{\ensuremath{\nu}}

\newcommand\U{{\ensuremath{\cal U}}}

\newcommand{\de}{\partial}
\newcommand{\ha}{\frac{1}{2}}

\newcommand{\ba}{\begin{eqnarray}}
\newcommand{\ea}{\end{eqnarray}}

\newcommand{\vel}{{\cal V}}

\usepackage{soul}

\makeatletter
\def\ps@mine{%
    \def\@oddfoot{\hfil\thepage\hfil}\let\@evenfoot\@oddfoot
    \let\@oddhead\@evenhead%
    \let\@mkboth\@gobbletwo
    \let\sectionmark\@gobble
    \let\subsectionmark\@gobble
    }
\renewcommand\section{\@startsection {section}{1}{\z@}%
                                   {-3.5ex \@plus -1ex \@minus -.2ex}%
                                   {2ex \@plus.2ex}%
                                   {\normalfont\large\sffamily\bfseries}}
\renewcommand\subsection{\@startsection {subsection}{1}{\z@}%
                                   {-3.5ex \@plus -1ex \@minus -.2ex}%
                                   {2ex \@plus.2ex}%
                                   {\normalfont\sffamily\bfseries}}
\makeatother

\usepackage{multirow}

\numberwithin{equation}{section}

\newcommand{\solid}{{elastic solid }}
\newcommand{\solids}{{elastic solids }}

\usepackage{cite}

\usepackage{setspace}

\allowdisplaybreaks

\begin{document}
\thispagestyle{empty}
\vspace*{-2.5cm}
\begin{minipage}{.45\linewidth}
\begin{flushleft}                           
{\footnotesize CERN-TH-2016-043}
\end{flushleft} 
\end{minipage}
\vspace{2.5cm}

\begin{center}
{\huge\sffamily\bfseries 
Massive and modified gravity \\\vspace{0.1cm} as self-gravitating media}
 \end{center}
 
 \vspace{0.5cm}
 
 \begin{center} 
{\sffamily\bfseries \large  Guillermo Ballesteros}$^{a,b}$,   {\sffamily\bfseries \large Denis Comelli}$^c$, {\sffamily\bfseries \large Luigi Pilo$^{d,e}$}\\[2ex]
  {\it
$^a$ Institut de Physique Th\'eorique, Universit\'e Paris Saclay, CEA, CNRS\\91191 Gif-sur-Yvette, France\\\vspace{0.1cm}
$^b$CERN, Theory Division, 1211 Geneva, Switzerland\\\vspace{0.1cm}
$^c$INFN, Sezione di Ferrara,  I-35131 Ferrara, Italy\\\vspace{0.1cm}
$^d$Dipartimento di Scienze Fisiche e Chimiche, Universit\`a di L'Aquila,  I-67010 L'Aquila, Italy\\\vspace{0.1cm}
$^e$INFN, Laboratori Nazionali del Gran Sasso, I-67010 Assergi, Italy\\\vspace{0.3cm}
{\tt guillermo.ballesteros@cea.fr}, {\tt comelli@fe.infn.it}, {\tt luigi.pilo@aquila.infn.it}
}
\end{center}

\vspace{0.7cm}

\begin{center}
{\small \today}
\end{center}

\vspace{0.7cm}

\begin{center}
{\sc Abstract}
\end{center}
\no

We study the effective field theory  that describes the low-energy physics of self-gravitating  media. The field content consists of four derivatively coupled scalar fields that can be identified with the internal comoving coordinates of the medium. Imposing SO(3) internal spatial invariance, the theory describes supersolids. Stronger symmetry requirements lead to superfluids, solids and perfect fluids, at lowest order in derivatives. In the unitary gauge, massive gravity emerges, being thus the result of a continuous medium propagating in spacetime. Our results can be used to explore systematically the effects and signatures of modifying gravity consistently at large distances. The dark sector is then described as a self-gravitating medium with dynamical and thermodynamic properties dictated by internal symmetries. These results indicate that the divide between dark energy and modified gravity, at large distance scales, is simply a gauge choice.

\newpage

\clearpage
\setcounter{page}{1}

\begin{spacing}{1}

\tableofcontents

\end{spacing}

\section{Introduction and outline}

The cosmological data show that the Universe is undergoing a phase of accelerated expansion. See \cite{SNe} for the first conclusive evidence. Although a simple model of the Universe based on a Friedmann-Lema\^itre-Robertson-Walker (FLRW) metric in  General Relativity (GR) supplemented by a cosmological constant  fits well all the observations \cite{PLANCK1,PLANCK2}, the actual nature of the driving force behind the  acceleration remains still unclear. Upcoming probes of the dynamics of the universe \cite{DE-probes}  will be crucial to shed light on this dark energy (DE). A good deal of  theoretical effort  has been aimed in the last twenty years to build consistent and compelling models of DE; see \cite{reviews} for reviews.\footnote{However, it is worth stressing that no such model solves the problem of the smallness of the cosmological constant \cite{Weinberg:1988cp}.} What the vast majority of them have in common is the addition of some new degrees of freedom to the dynamics of the universe. On the one hand, these hypothetical degrees of freedom have often been interpreted as extra components of the universe beyond baryonic and dark matter, photons and neutrinos. On the other hand, they have also  been commonly regarded as part of gravity itself, modifying the behaviour of GR at large distances, in a way that is compatible with the current acceleration of the Universe. 

In this work we propose a symmetry-driven  approach to DE. By working with an action of four derivatively coupled scalar fields, we show that DE can originate from a medium with concrete mechanical and thermodynamic properties. These four scalars are interpreted as comoving coordinates of the medium and, moreover, they are the degrees of freedom required to describe its low energy physics. At the core of this description of DE lie the symmetries of the medium. They rule the mutual interactions of the four scalars and determine whether the medium behaves as a perfect fluid, a superfluid, an elastic solid or a supersolid. 

By using this large class of Lagrangians, we will argue that the distinction that is often made between {DE}  (as some kind of matter) and modified gravity (as a true modification of Einsten's gravity) is purely artificial. Indeed both interpretations should be understood as complementary points of view of a unique (hypothetical) phenomenon. This fact becomes transparent by choosing the adequate coordinates for each perspective, as we explain below.

The four scalars can also be viewed as St\"uckelberg fields that allow to restore broken diffeomorphisms in models of so-called massive gravity  \cite{Dubovsky:2004sg,Rubakov:2008nh}. Following common usage, we use the term {\it massive gravity} (MG) to refer to any model of gravity for which non-derivative metric perturbations appear squared. We recall, however, that many such models do not contain a massive graviton (of spin two).  The introduction of  these fields is the key for interpreting massive (and general classes of modified) gravity models as a self-gravitating media. Thanks to diffeomorphism invariance, an adequate choice of the space-time coordinates (called unitary gauge) renders trivial the dynamics of the St\"uckelberg fields, allowing to draw a direct connection between the theory of self-gravitating media and MG.\footnote{An example of such media is the
  the case of the perfect fluids that can be described with three St\"uckelberg fields, for which the invariance under internal volume  preserving
  diffeomorphisms forbids a mass term for gravitons ($SO(3)$ spatial  tensor modes), see e.g.\ \cite{Dubovsky:2005xd,Ballesteros:2012kv}.} In particular, we describe in this work the matching to the general Hamiltonian construction of MG models that was studied in \cite{canonical-us,general-us,nonder-us}.
  
Most of such MG models are Lorentz breaking in the
sense that, around Minkowski spacetime, the metric perturbations are just
rotationally invariant in three dimensions.  
From the point of view of the self-gravitating medium interpretation of MG that we explore in this work, Lorentz breaking is simply triggered by
the presence of the medium itself, which generically defines at least one preferred four-vector.

By construction, the general description of self-gravitating media in terms of four derivatively coupled scalars is a well-defined effective field theory (EFT). This allows a systematic exploration of DE in terms of symmetries, with a unified and suggestive interpretation of modified and MG models. This is the take-home message of this paper. The use of an EFT framework to describe modified gravity at large distances an the acceleration of the universe, opens the possibility of confronting simultaneously a wide variety of models with cosmological observations in a simple way.\footnote{See also \cite{Gubitosi:2012hu} for an approach with a similar spirit, where only a single new degree of freedom is considered and non-derivative couplings are allowed.}

The outline of the paper  is the following. In Section \ref{Aseqm} we start introducing the EFT of self-gravitating media, emphasizing the role of diffeomorphism invariance. In Section \ref{Media} we continue the presentation of the EFT by building the intuition that allows to interpret the low-energy degrees of freedom as comoving coordinates. In Section \ref{sym} we write the lowest order scalars of the EFT according
to the possible symmetries. In Section \ref{Lagrange} we provide a discussion of the different kinds of media according to their symmetries and their field content. In Section \ref{massg} the unitary gauge is employed to show how the EFT of self-gravitating media can be put in correspondence with a broad class of MG models. In Section \ref{cosmox} we set the basis for a systematic study of the cosmology of these models by determining, for each kind of medium, the number of propagating degrees of freedom in a FLRW background. We present our conclusions in Section \ref{conc}. Two appendices are also included. In Appendix \ref{conse} we discuss briefly the conserved currents that appear in the theory.  Finally, in Appendix \ref{letus} we give an alternative, shortcut derivation of the number of propagating degrees of freedom.

Through the paper we will use the signature convention $(-,+,+,+)$ and units such that $\hbar = c =1$.

\section{Derivatively coupled scalar fields and diffeomorphism invariance} \label{Aseqm}
Let us consider the action for a set of four scalar fields under diffeomorphisms: $\Phi^A(x^\mu)$, $A=0,1,2,3$\,,  whose mutual interactions obey shift symmetries:
\be \label{ssyim}
\Phi^A \to \Phi^A + c^A \, , \qquad  \partial_\mu c^A =0 \,.
\ee
At lowest order in derivatives, the action for these fields depends on the kinetic blocks
\be \label{Bmatrix}
C^{AB} =g^{\mu \nu}  \de_\mu \Phi^A \de_\nu\Phi^B \,,
\ee
where $g_{\mu\nu}$ is the (four-dimensional) spacetime metric. Including the usual Einstein-Hilbert term the general form of the action at lowest order in derivatives is then given by
\be
S=  \int d^4x \, \sqrt{-g} \left[M_{pl}^2 \, R + U \left(C^{AB}\right) + L_{m} \right] \,,
\label{saction}
\ee
where we define $M_{pl}^2= 1/(16 \pi G)$, being $G$ Newton's constant;  $U$ is a smooth {\it master function} of the $4\times 4$ matrix $C^{AB}$ and  $L_{m}$ is the part of the
Lagrangian describing matter fields (e.g.\ baryonic matter).  The expression \eq{saction} is
the leading contribution to the action of an EFT where
shift symmetries enforce the fields $\Phi^A$ to be derivatively
coupled.\footnote{See \cite{Dubovsky:2005xd} for a general discussion of this type of EFT with an arbitrary number of scalar fields.}
The scalars $C^{AB}$ are the only independent ones that can be built out of the four $\Phi^A$ at
leading order (LO) in derivatives, i.e.\ with a single derivative
acting on each $\Phi^A$. The next-to-leading order in derivatives
(NLO) corresponds to operators that enter in the action with exactly two additional derivatives.\footnote{See \cite{Bhattacharya:2012zx,Ballesteros:2014sxa} for the NLO of a simpler case containing just three scalar fields.} These operators must then appear
suppressed with respect to the LO ones by some energy scale $\Lambda$ squared. In principle,
one would expect $\Lambda\sim U^{1/4}$, because this is the only scale of \eq{saction} other than $M_{pl}$ (and any masses coming from $L_m$).  

At LO, we are also 
allowed to multiply the curvature scalar $R$ by an arbitrary function of $C^{AB}$. However,
this function can always be eliminated through a conformal
transformation of the metric,\footnote{We follow here the nomenclature of \cite{Wald:1984rg}, but the reader must be aware that this is often called a {\it Weyl} transformation.} see \cite{Ballesteros:2014sxa}. After such a transformation, we obtain an action with the same structure of \eq{saction} plus some
extra terms which can be disregarded at LO because they are of higher order in
derivatives. Similarly, the Lagrangian $L_m$ can be multiplied by an
arbitrary function of $C^{AB}$. In fact, denoting by $\psi_m$ the
matter fields contained in $L_m$, according to the power-counting  this part of the action should actually be
written as $f(C^{AB})L_m[h(C^{AB})g_{\mu\nu},\psi_m]$, where $f$
and $h$ denote a set of arbitrary functions
\cite{Ballesteros:2014sxa}\,. In this paper we will not study the
effect of these couplings to matter fields, since our emphasis is on the interactions between the fields $\Phi^A$ and the metric $g_{\mu\nu}$. Therefore, in what follows we set $L_m$ to zero, postponing its study for future work.

Under these assumptions, the LO gravitational energy-momentum tensor (EMT) associated to the scalar fields $\Phi^A$, defined via the variation $\sqrt{-g}\,T_{\mu\nu}\delta g^{\mu\nu}=-2\delta\int d^4x \sqrt{-g}\, U$, is given by
\be \label{EMTS}
T_{\mu \nu} =   U  \, g_{\mu \nu} - 2 \, \frac{\de  U}{\de C^{AB}}  \,
\partial_\mu \Phi^A  \partial_\nu \Phi^B\,,
\ee
where sums over the repeated indices $A$ and $B$ are implicit. Varying the action \eq{saction}, the equations of motion (EOMs) of the scalars $\Phi^A$ read
\be
\nabla_\mu \left(\nabla^\mu \Phi^B\,\, {\de U}/{\de C^{AB}}  \right) =0 \,,
\label{gseqm}
\ee
where $\nabla_\mu$ is the covariant derivative defined with the
Levi-Civita connection of $g_{\mu\nu}$. Taking the covariant
divergence of \eq{EMTS}, and using that the Levi-Civita connection is
torsionless, we obtain
\begin{align} \label{emtc}
\nabla^\mu T_{\mu\nu}=\nabla_\nu\Phi^A\, \nabla_\mu\left(\nabla^\mu \Phi^B\,\, {\de U}/{\de C^{AB}} \right)\,.
\end{align}
Therefore, the covariant conservation of the EMT $(\nabla_\mu T^{\mu\nu}=0)$ can be inferred from the scalar's EOMs \eq{gseqm}. Clearly, this result would also hold if, $N$, the total number of scalars $\Phi^A$ were different from 4.

Generically, given some model of scalar fields coupled to gravity, the converse of the previous result is not true: the conservation of the EMT does not imply the EOMs of the scalars. However, in our case the two are equivalent under a specific condition. Let us consider \eq{saction} with $L_m=0$ and assume (as we will do in most of this paper) that we have four scalars $\Phi^A$. If  $\det(\partial_\mu \Phi^A) \neq 0$,  it is possible to choose a system of coordinates such that
\begin{align} \label{ntc}
\partial_\mu\Phi^A=\delta_\mu^A\,.
\end{align}
 Then, the equation \eq{emtc} becomes $\nabla^\mu T_{\mu\nu}=\delta_\nu^A\delta_\mu^B\, \nabla^\mu\left({\de U}/{\de C^{AB}} \right)$,
which is nothing but \eq{gseqm} and shows that the EOMs of the scalars now follow from the conservation of the EMT. Since $\nabla^\mu T_{\mu\nu}=0$ is ultimately due to diffeomorphism invariance,  we can say that the equations of motion are a consequence of it, provided that $\det(\partial_\mu \Phi^A) \neq 0$.\footnote{An analogous argument can be applied if the number of scalar fields, $N$, is smaller than the number of spacetime dimensions, $D$. However, if $N>D$ there would not be enough freedom to eliminate all the scalars from the dynamics through \eq{ntc} and so the EOMs cannot be equivalent the conservation of the EMT in that case.} A more direct way of proving it uses that \eq{emtc} can be expressed as the product $\mathcal{Q} \cdot q$, where $\mathcal{Q}$ is a $4\times 4$ matrix of components $\nabla_\nu\Phi^A$ and $q$ is a $4\times 1$ matrix. Then, if $\det \mathcal{Q}\neq 0$, the only solution of $\mathcal{Q} \cdot q=0$ is $q=0$, which is \eq{gseqm}. 

It is well-known that diffeomorphism invariance can be restored for any action of metric perturbations around some background. In four dimensions this can be done by introducing four scalars that are called St\"uckelberg fields.\footnote{The term {\it St\"uckelberg field} is commonly used for a field that allows to make explicit a (spontaneously broken) gauge symmetry. These fields and the associated mechanism (or ``trick"), which we will use later on, are named after E.C.G. St\"uckelberg, who showed how a (complex) scalar field can make an Abelian vector field massive while preserving gauge invariance  \cite{Stuck}. See \cite{Ruegg:2003ps} for a review with various applications.} By construction, these scalar fields are coupled among themselves only derivatively. From this point of view, the action \eq{saction} with $L_m=0$ is a good candidate for an EFT description of massive (and modified) gravity. The choice of $N=4$ scalar fields $\Phi^A$ is then motivated by the number of spacetime dimensions we assume. As the St\"uckelberg fields acquire a non trivial background configuration, such as e.g.\ \eq{ntc}, diffeomorphism invariance appears to be broken. 

MG models (like the archetypical
one of Fierz and Pauli \cite{Fierz:1939ix}) are often presented in a
way in which general covariance  is explicitly broken (and in some
cases Lorentz invariance too).  In  this work we exploit the fact that
these models can generically be embedded in a covariant (and Lorentz
invariant) action by the introduction  of four scalar St\"uckelberg fields. As we have just explained, the
apparent breaking of diffeomorphism invariance is just a consequence of choosing a background solution of
the equations of motion for these fields. From this perspective, the
action \eq{saction}, together with a
background satisfying  \eq{ntc}, is an appropriate tool for the description of
MG models, as we will discuss  later in more
detail.

The structure of \eq{saction} makes it convenient to assign dimensions of length to the fields $\Phi^A$. With this convention, the LO operators are dimensionless and the NLO ones have dimension of energy squared, which makes explicit the fact that they must appear suppressed by $\Lambda^2$. We can always express the St\"uckelberg fields $\Phi^A$ introducing new fields $\pi^A$ as follows:
\begin{align} \label{phonons}
\Phi^A(x^\mu) = x^A+\pi^A(x^\mu)\,.
\end{align}
By construction, the fields $\pi^A$ inherit the power-counting defined
for the St\"uckelberg fields $\Phi^A$. Inserting \eq{phonons} in  the
action \eq{saction},  assuming that we can neglect NLO and higher
order operators,  and imposing that $|\partial_\mu\pi^A|\ll 1$, we get
an action for the $\pi^A$ that we can expand in powers of
$\partial_\mu$ or, equivalently, in powers of $\pi^A$. With the
condition $|\partial_\mu\pi^A|\ll 1$, the field redefinition
\eq{phonons} suggests that the fields $\pi^A$ parametrize the
deviation of the fields $\Phi^A$ from their background location
$x^A$. Therefore, after the interpretation of $\Phi^A$ as coordinates
in a medium, the equation \eq{phonons} tells us that $\pi^A$ may be
understood as the degrees of freedom responsible for carrying sound waves in the
medium itself, i.e.\ phonons which can be understood as the Goldstone
bosons of broken translations, see e.g.\ \cite{Leutwyler:1993gf,Leutwyler:1996er}. We emphasize that these fields need not be small; it is only their derivatives that have to be under control for the validity of the derivative expansion. The fields $\pi^A$ are the four-dimensional generalization of the standard displacement field used in fluid dynamics, elasticity and strain theory. In the next section we start exploiting this connection, which we will phrase directly using the St\"uckelberg fields, using the language of the EFTs of solids, fluids and superfluids; different aspects of which have been  studied in \cite{Dubovsky:2005xd,Endlich:2010hf, Dubovsky:2011sk,Dubovsky:2011sj,Nicolis:2011cs,Hoyos:2012dh,Endlich:2012pz,Endlich:2012vt,Ballesteros:2012kv,
Bhattacharya:2012zx,Ballesteros:2013nwa,Nicolis:2013lma,Ballesteros:2014sxa,Gripaios:2014yha,Delacretaz:2014jka}. As a matter of fact, the basic idea of using three comoving coordinates to describe fluid dynamics is nothing but the earlier {\it pull-back formalism}, which had already been used to describe the dynamics of continuous media,  see e.g.\ \cite{pullb,Carterbook,Brown:1992kc, Comer:1993zfa, Comer:1994tw, Andersson:2006nr}.   

\section{Self-gravitating media} \label{Media}
We split the scalars $\Phi^A$ into two distinct sets: $\{\Phi^0\}$ and $\{\Phi^1,\Phi^2,\Phi^3\}$, which are characterized as follows. The elements of $\{\Phi^1,\Phi^2,\Phi^3\}$, will be referred to as {\it spatial} St\"uckelberg fields, and we will use small latin letters $a,b, \cdots$ to index them. 
We require the $3\times 3$ submatrix of \eq{Bmatrix} whose components are
\be \label{post}
B^{ab}\equiv C^{ab}
\ee
to be positive-definite, so that its three eigenvalues are positive. As we will soon see, the reason for this condition is allowing the inverse of $B^{ab}$ to define an Euclidean metric in $\myR^3$. As anticipated in the previous sections, we can interpret the fields $\Phi^a$ as the (spatial) comoving coordinates of an internal space  ${\cal F}$ that describes
a continuum medium whose topology we assume to be $\myR^3$. This
picture --that has been known for a long  time, see e.g.\ \cite{Carterbook}--
allows to write an unconstrained variation principle for a 
fluid. A pedagogical review of this {\it pull-back formalism} is provided in \cite{Andersson:2006nr} and various applications specific to cosmology can be found in \cite{Gruzinov:2004ty,Blas:2012vn,Endlich:2012pz,Ballesteros:2012kv,Ballesteros:2013nwa,Bartolo:2013msa,Endlich:2013jia,
Ballesteros:2014sxa,Skordis:2015yra,Berezhiani:2015bqa}. Here, we will follow \cite{Ballesteros:2012kv} to describe the basics of this construction. 

We exploit the fact  that, given a slicing and a threading of the  spacetime (Lorentzian) manifold $\cal M$, we can formally write $\cal M=\cal{M_T}\times\cal{M_S} $, where $\cal{M_T}$ and $\cal{M_S}$ are respectively the sets of time, $\{x^0\}$, and space, $\{x^1,x^2,x^3\}$, coordinates on $\cal{M}$. By choice, $\cal M_S$ is endowed with a Riemannian metric. Provided that at any given time $x^0$, the condition $\det(\partial_i \Phi^a)\neq 0$ is satisfied, there exists an invertible map of ${\cal M_S}$ into the medium ${\cal F}$, which we assume to be smooth: $x^i\rightarrow \Phi^a(x^0,x^j)$. Conversely, ${\cal F}$ can be immersed in ${\cal M_S}$ via the inverse map: $\Phi^a\rightarrow x^i(x^0,\Phi^a)$. While the first of these maps  identifies the fluid (or in general, medium) element (labelled by $\Phi^a$) that sits at time $x^0$ at the point $x^i$; the second map describes the trajectory of the medium element $\Phi^a$ in spacetime \cite{Ballesteros:2012kv}. This is nothing but the dual description of a continuum in terms of Lagrangian ($\Phi^a$) and Eulerian ($x^i$) coordinates. 

The spacetime metric $g_{\mu\nu}$ induces a three-dimensional metric on  ${\cal F}$, which is given by the matrix $B_{ab}$ defined as follows:
\be \label{rmet}
ds^2_{ {\cal F}} = g_{\mu \nu} \frac{\de x^\mu}{\de \Phi^a} \frac{\de
  x^\nu}{\de \Phi^b} \, d\Phi^a d\Phi^b \equiv B_{ab}  \, d\Phi^a d\Phi^b \,.
\ee
The induced metric $B_{ab}$ is the inverse matrix of $B^{ab}$ defined in \eq{post}, i.e.\ $B^{ab}B_{bc} = \delta^a_c$. The condition that $B^{ab}$ is positive definite ensures that the induced spatial metric on the medium $\cal F$ is Riemannian. 

Since the fields $\Phi^a$ can be interpreted as comoving coordinates, they must remain unchanged along the fluid flow. This allows to define a (unique) four-velocity $u^\mu$, through the conditions
\be
u^\mu \, \de_\mu \Phi^a =0 \, , \qquad u^\mu u^\nu g_{\mu \nu} =-1\,,
\label{fvel}
\ee
whose only solution is
\be
u^\mu = -\frac{\epsilon^{\mu \nu \alpha \beta}}{6 \, b\sqrt{-g} }\epsilon_{abc}
\, \de_\nu \Phi^a \, \, \de_\alpha \Phi^b \, \de_\beta \Phi^c \,
 \,,
\label{veloc}
\ee
where
\be \label{detb}
\qquad b \equiv\sqrt{\det\,\pmb B}\,,
\ee
and $\pmb B$ denotes the $3\times 3$ matrix whose components are given by \eq{post}. In what follows, we will always use boldface capital letters for three-dimensional spatial matrices.

When pulled-back into the spacetime manifold $\mathcal M$, the induced metric $B_{ab}$ becomes a projector $H_{\mu \nu}$
on the embedding  of ${\cal F}$ in ${\cal M}$:
\be 
H_{\mu \nu} = B_{ab} \de_\mu \Phi^a \, \de_\nu \Phi^b
\equiv g_{\mu \nu} + u_\mu \, u_\nu \, , \qquad H_{\mu \nu} u^\nu=0 \,.
\label{ind}
\ee
Only if $u_{[\alpha} \nabla_\mu u_{\nu]}  =0 $ (where the brackets indicate full antisymetrization of the three indices), Frobenius theorem guarantees that  $u^\mu$ is hypersurface orthogonal. In that case there is a one-parameter family of hypersurfaces that are orthogonal (everywhere) to the four-velocity $u^\mu$, which thus defines a slicing of the spacetime manifold $\cal M$.

So far we have only dealt with the spatial St\"uckelberg fields, describing the structure that was used, e.g.\ in \cite{Ballesteros:2012kv} to study (a certain class of) perfect fluids. We will now deal as well, with the {\it temporal} St\"uckelberg, $\Phi^0$, for which we impose the condition
\be \label{c00}
X\equiv C^{00} < 0\,.
\ee
This condition allows to define another four-velocity by means of the derivative $\partial_\mu\Phi^0$, which is thus constrained to be time-like: 
\be \label{vdef} 
\vel^\mu = -\frac{\partial^\mu \Phi^0}{\sqrt{{-X}}} \, , \qquad\vel^\mu\vel^\nu g_{\mu\nu} =-1 \,.
\ee
This defines a hypersurface at each point that is orthogonal to $\vel^\mu$ and has an induced metric
\be
p_{\mu \nu} = g_{\mu \nu} + \vel_\mu \, \vel_\nu \,,\qquad p_{\mu\nu} \vel^\nu =0\,.
\ee
Since $\vel_\mu$ is the gradient of a scalar, its covariant vorticity $\omega_{\mu\nu}= p^\alpha_\mu \,  p^\beta_\nu \, \nabla_{
 \alpha } \vel_{\beta} $ vanishes. In Minkowski and FLRW spacetimes, the four-vectors $u^\mu$ and $\vel^\mu$ 
coincide, up to a sign. However, if these spaces are slightly distorted, the two velocities generically differ from each other. Given the four scalar fields $\Phi^A$, the four-vectors $\vel^\mu$ and $u^\mu$ are the only independent four-vectors that are invariant under the group $SO(3)_s$ of internal spatial rotations and under the shift symmetries \eq{ssyim}. 

The condition \eq{c00} also allows an interesting extension of the maps between $\cal M_S$ and $\cal F$ that we mentioned earlier. In particular we can extend the medium space $\cal F$, spanned by the three spatial St\"uckelberg fields $\Phi^a$, to a space ${\cal F}_4$ spanned by the full set $\{\Phi^0\,,\Phi^a\}$. By construction, and in analogy to \eq{rmet},  the matrix $C_{AB}$, defined through
\begin{align}
C_{AB}\equiv g_{\mu\nu}\frac{\partial x^\mu}{\partial \Phi^A}\frac{\partial x^\nu}{\partial\Phi^B}\,,
\end{align}
is a Lorentzian metric in the extended medium space $\mathcal F_4$; and is the inverse of the $4\times 4$ matrix of kinetic blocks introduced in \eq{Bmatrix}. Hence, the map
\be
 {\cal M} \to  {\cal F}_4: \, x^\mu \to \Phi^A(x^\mu)  
\label{sub}
\ee
carries the spacetime manifold onto the extended (self-gravitating) medium space, whereas the inverse 
\be \label{sub2}
 {\cal F}_4 \to  {\cal M}: \, \Phi^A \to x^\mu(\Phi^A)
\ee
gives the location of each element of the extended medium (labelled by $\Phi^A$) on spacetime, as described by the coordinates $x^\mu$. In other words, the second of these two maps describes the world hyper-volume of the extended medium ${\cal F}_4$ as it propagates on the spacetime $\cal M$. 

The role of the temporal St\"uckelberg $\Phi^0$ is therefore two-fold. First, allows to define an independent time-like four-vector via \eq{vdef}. As we will later see, this will give us the possibility of defining two-constituent fluids, as required to describe superfluids (provided that the appropriate symmetries are satisfied) \cite{Nicolis:2011cs}. And second, as we have just described, $\Phi^0$ can also be naturally  interpreted as a comoving time coordinate in the medium space, which allows to conceive the self-gravitating media described by \eq{saction} as four-dimensional objects propagating in spacetime, extending the pull-back formalism to include the temporal dimension as well. Both pictures play important roles in the novel interpretation of MG and DE as self-gravitating media that we explore in this work. 

In the next section we focus on the operators that can appear in the action \eq{saction} at LO in derivatives, and in Section \ref{Lagrange} we will classify the different media that arise combining them. 

 \section{Spatial SO(3) invariance and additional symmetries} 
\label{sym}
For simplicity, we will always impose  a global internal spatial $SO(3)$ symmetry in the medium space, that we denote by $SO(3)_s$,  so that all the operators in the action are invariant
under 
\be \label{rots}
 \Phi^a\to \;R^{ab}\;\Phi^b \qquad \text{with} \qquad \pmb R \in SO(3)\,,\quad \partial_\mu \pmb R =0\,.
\ee
This basic symmetry will also allow us to make direct contact with the general models of massive gravity classified in \cite{nonder-us}, as we will discuss in Section \ref{massg}. 

To construct a LO complete set of independent $SO(3)_s$-invariant scalars we start with the traces
\begin{align}
\tau_n =\text{Tr}({\pmb B}^n)\,,\quad n=1,2,3\,. \label{traces}
\end{align}
We recall that the $SO(3)_s$-invariant determinant $b=\sqrt{\text{det}\,{\pmb B}}$, defined already in \eq{detb}, is not independent of these traces because it can be written as $b^2=(\tau_1^3-3\,\tau_1\,\tau_2+2\,\tau_3)/6$.

Introducing the $3\times 3$ matrix $\pmb Z$ of components
\be
Z^{ab} = C^{0a}C^{0b}\,,
\ee
we can define the scalars
\begin{align} \label{defy}
y_n & = \text{Tr} \left({\pmb B}^n \cdot \pmb{Z} \right)\,, \quad n=0,1,2,3\,,
\end{align}
which are also invariant under $SO(3)_s$.  The traces of $\pmb Z^2$ and $\pmb Z^3$ (and higher powers) do not need to be considered once $y_0$ is included because they can be written as powers of the latter.   

Besides, since the four-vectors $ \partial_\mu \Phi^0$ and $u_\mu$ are invariant under $SO(3)_s$, there are two other scalars at LO  that are invariant under this group. One of them is $X=C^{00}$, which we already introduced in \eq{c00}, and the other is 
\be \label{4ddet}
Y \equiv u^{\mu}\partial_\mu\,\Phi^0\,.
\ee
This closes the set of independent LO $SO(3)_s$-invariant operators, which is therefore composed of nine elements and reads: 
\begin{align} \label{ooops}
\mathcal{O}_{\text{LO}}=\{X \, , Y \,,\tau_1\,,\tau_2\,,\tau_3\,,y_0\,,y_1\,,y_2\,, y_3\}\,.
\end{align}

We point out that the determinant of the $4\times 4$ matrix of components $C^{AB}$, defined in \eq{Bmatrix}, is also invariant under $SO(3)_s$, but it is not independent of this set because it can be expressed as
\be 
\sqrt{|\det(C^{AB})|}=Y \,b\, , 
\ee
What is interesting of this operator is that is closely related to the pull-back of the medium volume element in
spacetime. Indeed
\be \label{vicev}
 d^4 \Phi = \sqrt{-g} \, b \, Y  \, d^4x \,.
\ee
This expression allows to pass from Eulerian to Lagrangian coordinates (and viceversa) in four dimensions, extending the three-dimensional treatment of these coordinates that was discussed in \cite{Ballesteros:2012kv}. 

\subsection{Additional symmetries at LO} \label{wecf}

\begin{table}[t]
  \footnotesize
  \small
  \renewcommand\arraystretch{1.5}
  \centering
  \begin{tabular}{|c|c|}
    \hline{\bf Operator} &{\bf Definition}    \\ 
    \hline
    \hline
$C^{AB}$ & $g^{\mu \nu} \, \de_\mu \Phi^A \, \de_\nu \Phi^B$\,,\quad \scriptsize{$A,B=0,1,2,3$}\\  \hline 
$B^{a b}$ & $g^{\mu \nu}\partial_\mu \Phi^a \, \partial_\mu \Phi^b$\,,\quad \scriptsize{$a,b=1,2,3$}\\ \hline 
$Z^{ab}$ & $C^{a0} \, C^{b0}$ \\ \hline
$X$  & $C^{00}$\\ \hline
$W^{ab}$ & $B^{ab} - Z^{ab}/X$\\ \hline
 $b$ &  $\sqrt{\det \pmb{B}}$ \\ \hline
  $Y$    & $u^\mu \partial_\mu \Phi^0$ \\ \hline
   $y_n$ & $\text{Tr} \left({\pmb B}^n \cdot \pmb{Z} \right)$\,, \quad \scriptsize{$n=0,1,2,3$}\\ \hline
$\tau_n$ & $\text{Tr}\left(\pmb{B}^n \right)$\,, \quad \scriptsize{$n=1,2,3$}\\ \hline
$w_n$ & $\text{Tr}\left(\pmb{W}^n \right)$\,, \quad \scriptsize{$n=0,1,2,3$}\\ \hline
${\cal O}_{\alpha\beta n}$ &  $\left({X}/{Y^2}\right)^\alpha \, \left({y_n}/{Y^2}\right)^\beta$\,,\quad \scriptsize{$\alpha\,,\beta \in \myR $} \\ \hline
${\cal O}_{\alpha}$ &  $\left({X}/{Y^2}\right)^\alpha$\,,\quad \scriptsize{$\alpha \in \myR $}  \\ \hline

       \end{tabular}
  \caption{\it Summary of the various LO scalar operators appearing in this paper. Unless it is explicitly indicated otherwise, we use Greek letters for spacetime indices, capital Latin letters  {\scriptsize ($A,B,\ldots=0,1,2,3$)}  for indices in the internal spacetime of the medium and small Latin letters {\scriptsize ($a,b,\ldots=1,2,3$)} for spatial indices of the medium. Boldface Latin capital letters, such as $\pmb{B}$,  represent $3\times 3$ matrices.}
  \label{tab:operators}
\end{table}

We can further constrain the form of the action \eq{saction} by requiring additional symmetries, leading to different types of self-gravitating media. Two examples are \solids and perfect fluids, which are derived from symmetries imposed on the spatial sector of the St\"uckelberg fields. We will see in Section \ref{Lagrange} that by imposing symmetries that mix the spatial and temporal St\"uckelberg fields, new kinds of materials arise. For reference, we summarize in Table \ref{tab:operators} the notation for the operators that we use in this paper. Let us now make a list of the additional symmetries of interest for us:

\begin{itemize}
\item  Imposing volume-preserving diffeomorphism invariance of the spatial sector: 
\be 
V_s\text{Diff}:\quad\Phi^a\to \Psi^a(\Phi^b)\,,\quad \det\left(\frac{\partial\Psi^a}{\partial\Phi^b}\right)=1\,,
\label{mgauge}
\ee
the scalars $\tau_n$ have to combine in the action into the determinant operator $b$ and, in addition, the scalars $y_n$ are automatically forbidden. Therefore $b$, $X$ and $Y$ are the only operators allowed at LO. This symmetry was identified in \cite{Dubovsky:2005xd} in the context of the EFT formulation of perfect fluids containing only $b$. 
\end{itemize}

\noindent We will also consider a series of symmetries that were proposed in \cite{Dubovsky:2004sg}:

\begin{itemize}
\item 
If we impose that the action has to be invariant
  under the transformations 
\be \label{mgaugezero}
\Phi^a \to \Phi^a + f^a(\Phi^0)\,,
\ee
for arbitrary $f^a$, we find that at LO the  action will generically depend only on $X$ and on the matrix of 
components
\be
W^{ab} \equiv  B^{ab} - \frac{Z^{ab}}{X}\,.
\ee
Then, the invariance under $SO(3)_s$ further enforces $W^{ab}$ to appear in the action only through the traces
\begin{align}
w_n\equiv \text{Tr}(\mathbf{W}^n)\,,
\end{align}
which are non-linear combinations of the scalars $X$, $\tau_n$ and $y_n$:
\be\label{Wn}
w_1=\,\tau_1-\frac{y_0}{X},\quad
w_2=\,\tau_2-2\,\frac{y_1}{X}+\frac{y_0^2}{X^2},\quad
w_3=\,\tau_3-3\,\frac{y_2}{X}+3\,\frac{y_0\,y_1}{X^2}-\frac{y_0^3}{X^3}.
\ee 

Under the transformation \eq{mgaugezero}, $u^\mu \de_\mu \Phi^a  \to \ u^\mu \de_\mu \Phi^a  + Y \, 
{df^a}/{d\Phi^0}$, because $u^\mu$ is invariant. Therefore, the operator $Y$ measures the failure of $\Phi^a$ to remain comoving along
$u^\mu$ when such a transformation is applied. Besides, as in the case of $V_s$Diff, at NLO the same operators with derivatives of $u^\mu$ are also allowed under \eq{mgaugezero}.
\item
If the action is invariant under 
   \be
\Phi^0 \to \Phi^0 + f(\Phi^a) \,,
\label{stt}
\ee
the only LO operators that are allowed are $Y$ and the traces $\tau_n$. Clearly, $u^\mu$ is invariant under this transformation and therefore all the NLO operators mentioned above constructed from derivatives fo $u^\mu$ also respect this transformation. 

If we combine the symmetries \eq{stt} and \eq{mgauge} the master function $U$ can only depend on $Y$ and $b$ and the resulting medium has the EMT of a perfect fluid, as we will later see. 
 
\item  Finally, if the action is invariant under
\be
\Phi^0 \to \Phi^0 + f(\Phi^0) \,,
\label{stt2}
\ee
it is constrained at LO to be a function of the traces $\tau_n$, $w_n$ and the scalars
\be \label{Oabn}
{\cal O}_{\alpha \beta n} = \left(\frac{X}{Y^2}\right)^\alpha \, \left(\frac{y_n}{Y^2}\right)^\beta \, .
\ee
The operators ${\cal O}_{\alpha \beta n}$ present some peculiarities with respect to the ones that we have encountered so far. From the point of view of the symmetries, $\alpha,\beta\in\myR$, which implies that there must be an infinite (and uncountable) amount of these operators.  However, since all $y_n$ vanish at zero order for diagonal metric backgrounds, a well defined perturbative
expansion seems to require $\beta$  to be a positive integer. Even then, there is still an infinite number of these operators at LO, which means that the master function $U$ cannot be arbitrary if the action is to be finite. The same difficulty is also manifest on the associated Noether currents, see Appendix  \ref{noth}.

If in addition to \eq{stt2}, we impose also \eq{mgauge}, the operators \eq{Oabn} get restricted to the subset with $\beta=0$. For later convenience we will denote them as follows:
\begin{align}\label{On}
\mathcal{O}_\alpha=\left(\frac{X}{Y^2}\right)^\alpha\,.
\end{align}

If instead we combine \eq{stt2} and \eq{mgaugezero}, the operators $w_n$ get selected. 

\end{itemize}
 
 In Table \ref{tab:summa} we summarize the different LO scalar operator content according to the internal symmetries of the action and the physical interpretation of the resulting system, which we discuss in the next section. In addition, media with reduced internal dimensionality are possible as well; see Table \ref{3dimmedia}. For instance, if the spatial St\"uckelberg fields $\Phi^a$ are not present in the LO action, the only operator at that order would be $X$. 

\begin{table}[t]
  \footnotesize
  \small
  \renewcommand\arraystretch{1.5}
  \centering
  \begin{tabular}{|c|c|c|}
 \hline
   \multicolumn{3}{|c|}{\bf Four-dimensional media}\\
    \hline
    \hline
{\bf Symmetries of the action} &{\bf LO scalar operators} &  {\bf Type of medium}    \\ 
    \hline
    \hline
$SO(3)_s$\quad \&\quad $\Phi^A\to\Phi^A+f^A$\,,\quad $\partial_\mu f^A=0$ &   $X$, $Y$, $\tau_n$, $y_n$ &  \multirow{5}{*}{supersolids}  \\ \cline{1-2}
  $\Phi^a\to \Phi^a+f^a(\Phi^0)$   &  $X$, $w_n$ & \\ \cline{1-2}
  $\Phi^0\to \Phi^0+f(\Phi^a)$   & $Y$, $\tau_n$ &  \\\cline{1-2}
$\Phi^0\to \Phi^0+f(\Phi^0)$   & $\tau_n$, $w_n$, $\mathcal{O}_{\alpha\beta n}$ &  \\ \cline{1-2}
  $\Phi^a\to \Phi^a+f^a(\Phi^0)$\, \&\, $\Phi^0\to \Phi^0+f(\Phi^0)$  &  $w_n$ &  \\ \hline
$V_s$Diff: $\Phi^a\to\Psi^a(\Phi^b)$\,,\quad$\det |\partial \Psi^a /\partial{\Phi^b}|=1$  &  $b$, $Y$, $X$ & \multirow{2}{*}{superfluids} \\\cline{1-2}
        $\Phi^0\to \Phi^0+f(\Phi^0)$ \, \&\, $V_s$Diff  &  $b$, $\mathcal{O}_{\alpha}$ &  \\ \hline
    $\Phi^0\to \Phi^0+f(\Phi^a)$ \, \&\, $V_s$Diff  &  $b$, $Y$ & perfect fluid \\ \hline
$\Phi^A\to\Psi^A(\Phi^B)$\,,\quad$\det |\partial \Psi^A 
       /\partial{\Phi^B}|=1$  &  $b\, Y$ &  perfect fluid with $\rho+p=0$ \\\hline 
       \end{tabular}
  \caption{\it Summary of local symmetries in material spacetime and
    the corresponding invariant scalar operators. Invariance under $SO(3)_s$ and shift symmetries are assumed by default in all cases.}
  \label{tab:summa}
\end{table}

\begin{table}[t]
  \footnotesize
  \small
  \renewcommand\arraystretch{1.5}
  \centering
  \begin{tabular}{|c|c|c|}
 \hline
   \multicolumn{3}{|c|}{\bf Media with reduced internal dimensionality}\\
    \hline
    \hline
 {\bf Symmetries of the action} &{\bf LO scalar operators} & {\bf Type of medium}    \\ 
    \hline
    \hline
     $SO(3)_s$\quad \&\quad $\Phi^a\rightarrow \Phi^a+c^a\,,\,\, \partial_\mu c^a =0$ &  $\tau_n$ & solid \\\hline
 $V_s$Diff & $b$ & \multirow{2}{*}{perfect fluid}\\ \cline{1-2}
$\Phi^0\rightarrow \Phi^0+c^0\,,\,\, \partial_\mu c^0 =0$ & $X$ &\\ \hline
       \end{tabular}
  \caption{\it Summary of self-gravitating media with reduced dimensionality, i.e.\ with less than four St\"uckelberg fields.}
  \label{3dimmedia}
\end{table}

\section{Lagrangians for self-gravitating media at LO} \label{Lagrange}
Given the field content: the metric $g_{\mu \nu}$ and the scalar 
fields $\Phi^A$, it  is possible to study
in a systematic way the  models resulting from imposing certain symmetries in addition to $SO(3)_s$ and internal shifts. 
Each class of symmetries corresponds to a medium with specific mechanical and thermodynamic properties, which
can be obtained from the EMT. For all the media we consider we require spacetime
diffeomorphism invariance. Then, the most general action at LO for a medium described by the four St\"uckelberg fields $\Phi^A$ can be constructed in
terms of the scalar invariants $X$, $Y$, $\tau_n$ and $y_n$:
 \be
\label{LAll}
 S=M_{pl}\int d^4 x\,\sqrt{-g}\, R+\int d^4x\sqrt{-g} \, U(X, \, Y, \,  \tau_n, \, y_n)
 \ee
and the corresponding EMT is given by
\be
 T_{\mu\nu}=U\,g_{\mu\nu} -2 \frac{\de U}{\de C^{AB}}\de_\mu \Phi^A \, \de_\nu \Phi^B=
U\, g_{\mu\nu}-
2\,\sum_k U_{\mathcal{O}_k}\, \frac{\partial \mathcal{O}_k}{\partial
  g^{\mu\nu}}\,,
 \ee
where $\mathcal{O}_k$ are the nine scalar LO operators appearing in \eq{LAll} and we use the notation $U_{\mathcal{O}_k}=\partial U/\partial \mathcal{O}_k$.  Their partial derivatives with respect to the inverse spacetime metric are 
\begin{equation} \label{expsk}
\begin{aligned}
\frac{\partial Y}{\partial g^{\mu\nu}}=-\frac{Y}{2}\;u_\mu\,u_\nu\,,\quad
  \frac{\partial X}{\partial g^{\mu\nu}}= -X\,\vel_\mu \vel_\nu\,,\quad
   \frac{\partial \tau_n}{\partial g^{\mu\nu}}=n\,\,\partial_\mu\Phi\cdot\boldsymbol{B}^{n-1}\cdot\partial_\nu\Phi\,,\quad \frac{\partial b}{\partial
  g^{\mu\nu}}=\frac{b}{2}\;H_{\mu\nu} \,, \\
   \frac{\partial y_n}{\partial g^{\mu\nu}}=\sum_{m=1}^{n}\partial_\mu\Phi\cdot\boldsymbol{B}^{n-m}\cdot\boldsymbol{Z}\cdot\boldsymbol{B}^{m-1}\cdot \partial_\nu\Phi
   {-2\sqrt{-X}\left(\mathcal{C}\cdot \boldsymbol{B}^n\cdot \partial_{(\mu} \Phi\right) \vel_{\nu)}}\,,   \quad\quad\quad\quad\quad
\end{aligned}
\end{equation}
where {the indices $\mu$ and $\nu$ are symmetrized in the second term of the last expression. We  have included in \eq{expsk} the operator} $b$, that can be written as a combination of the three $\tau_n$, for later convenience. The dot ($\cdot$) represents the standard three-dimensional matrix product and we have introduced the notation 
\begin{align}
\mathcal{C}^i = C^{0i}.
\end{align} 
In the expressions \eq{expsk},  $\mathcal{C}$ and $\partial_\mu\Phi$ have to be understood as a $1\times 3$ matrices of components $\mathcal{C}^i$ and $\partial_\mu\Phi^i$, respectively. It is also useful to note that 
\be
\frac{\partial u^\alpha}{\partial g^{\mu\nu}}=-\frac{u^\alpha}{2}\;u_\mu\,u_\nu\,,\quad \frac{\partial\mathcal \vel^\mu}{\partial g^{\alpha\beta}}=-\frac{\vel^\mu}{2}\vel_\alpha\vel_\beta\,.
\ee

In what follows we list the media that arise according to the symmetries that we have discussed and, also, according to the assumptions on the low-energy field content. We focus first on the systems that contain only a partial subset of the St\"uckelberg fields, which we call {\it temporal} or {\it spatial media}. These are the solids $U(\tau_n)$, and two different types of perfect fluids: $U(b)$ and $U(X)$; the first of which is a special case of the solids. Then we will move on to the media that at low-energy require all the St\"uckelberg fields. As discussed below, the properties of the most generic class of them have been argued to define (non-relativistic) supersolids \cite{Son:2005ak}. A subclass of these (with an enhanced symmetry group: internal three-volume preserving diffeomorphisms $V_s$Diff) would then describe superfluids. Other media, defined by specific additional symmetries,  will also be mentioned, see Table \ref{tab:summa}. The connection with modified and, specially, massive gravity models will be studied in Section \ref{massg}. Then, in Section \ref{cosmox} we will discuss the number of propagating degrees of freedom in FLRW, which strongly depends on the symmetries beyond \eq{ssyim} and \eq{rots} that are imposed.  

\subsection{Media with reduced internal dimensionality}

We start by describing three special cases for which some of the four
St\"uckelberg fields are missing from the LO action, while the shifts symmetries (and eventually while $SO(3)_s$)
are respected. We refer to these media as
{\it temporal} or {\it spatial} depending on whether the missing
fields are the spatial or temporal St\"uckelbergs,
respectively. Concretely, we will discuss two kinds of spatial media
(solids and perfect fluids) and the only kind of temporal medium  (an
irrotational perfect fluid) that exists at LO with our symmetry assumptions. A
peculiarity of the spatial and temporal media is that
there is no symmetry that forbids one subset of St\"uckelberg fields
at LO, but allows them at a higher order in derivatives. This means
that the defining restriction on the low-energy field content of these
models must be taken as an assumption, which in principle cannot be
justified from the point of view of the symmetries. 
 
\subsubsection{${U(\tau_1,\tau_2,\tau_3)}$: solid } \label{ssolid}

Let us first consider the possibility  that the only  operators relevant at LO are the traces $\tau_n$. Then, the action contains only the three spatial St\"uckelberg fields $\Phi^a$. This necessarily assumes that no other fields are needed to describe the system at sufficiently low energies. As we just mentioned, there is no symmetry that forbids $\Phi^0$ at LO but allows it at some higher order in derivatives. Notice that by looking at Table \ref{tab:summa}, it would seem that combining $\Phi^0\rightarrow \Phi^0+f(\Phi^0)$ and $\Phi^0\rightarrow \Phi^0+f(\Phi^a)$, the operators $\tau_n$ are selected. However, it is clear that this cannot be considered a true symmetry of an action based on $U(\tau_n)$ since $\Phi^0$ is not part of $\tau_n$.

In addition, given that the general action $\eq{LAll}$ lacks potential terms and it generically mixes all fields in a democratic way, it is difficult to imagine how $\Phi^0$ could be (on average) frozen at some constant value only at LO, if it is not neglected also at higher orders.  Therefore, it seems that if $\Phi^0$ does not appear at LO, it should be completely absent from the action and this must be imposed as an assumption on the physical nature of the system.   

The {\it solid} is thus the most general medium that we can construct at LO with only the three spatial St\"uckelberg fields, respecting the basic symmetries that we assumed ($SO(3)_s$ and shift invariance). The corresponding EMT can be written as
\be
T_{\mu \nu} = U \,  g_{\mu \nu}  -2\, U_{ab}\,\de_\mu \Phi^a \de_\nu
\Phi^b\,,
\label{tela}
\ee
where
\begin{align}
U_{ab}=U_{\tau_1} \, \delta^{ab} + 2 
\, U_{\tau_2} \, B^{ab} + 3 \, U_{\tau_3} \,  B^{ac} \, B^{cb}\,.
\end{align}
Such a medium can be interpreted as the relativistic generalization of  an elastic material, in the sense that this kind of media can support compressing pressures; see e.g.\ \cite{elastic}.
 In standard elasticity theory  in a flat space spacetime \cite{Landau}, the stress state is described by the spatial internal inverse metric $B^{ab}$, so that the medium is relaxed (not stressed) if $B^{ab} =\delta^{ab}$.
  
  The object $B^{ab}$ is the covariant generalization of the standard three-dimensional  left Cauchy-Green deformation tensor; and thus $\partial\Phi^a/\partial x^i$ is the deformation gradient tensor, see e.g.\ \cite{bower2009applied}.  This type of medium, which has been commonly called solid --see e.g.\ \cite{Leutwyler:1996er}-- or \solid --to distinguish it from a perfect fluid-- has been applied in astrophysics and cosmology for  the description of the dynamics of the interior of neutron
stars \cite{Carter-neutron}, in a proposal for a dark matter model \cite{Bucher:1998mh} and also for constructing a model of primordial inflation \cite{Gruzinov:2004ty,Endlich:2012pz}. 

Solids posses two features, due to the independent operators $\tau_n$, that make them different from perfect fluids --which we discuss below-- and interesting for certain  applications in the context of cosmology. First, they exhibits anisotropic stress, due to the pressure perturbations being in general dependent on the spatial direction. And second, the spatial trace operators $\tau_n$ are the only ones among the nine LO scalars operators \eq{ooops} that generate a mass for the (spin two) graviton, as we will see in Section \ref{massg}. Clearly, these two properties are not exclusive of the solids with EMT \eq{tela} --see Table \ref{tab:summa}-- but these are the simplest systems that display them. 

Although it is not immediate to identify a four-velocity from the EMT \eq{tela}, the natural choice is $u^\mu$, defined in \eq{veloc}, since this is the only $SO(3)_s$-invariant four-vector that is available when only the three spatial St\"uckelberg fields are present. With this choice of frame, the energy density and the pressure of the solid, defined as usual through the projections $\rho =T^{\mu\nu}u_\mu u_\nu$ and $3p=T^{\mu\nu}H_{\mu\nu}=T^{\mu\nu}(g_{\mu\nu}+u_\mu u_\nu)$,  are given by
\begin{align}
\rho = -U\,,\quad p=U- \frac{2}{3} \sum_{n=1}^3 n\, U_{\tau_n}\tau_n\,.
\end{align}
The four-velocity $u^\mu$ corresponds to the energy frame --sometimes called rest frame-- of the solid, which for an arbitrary EMT is defined (if it exists) as the four-vector $U^\mu_{(r)}$ that solves $\left(g^{\mu\nu}+U^\mu_{(r)} U^\nu_{(r)}\right) U^\gamma_{(r)} T_{\nu\gamma}=0$, see e.g.\ \cite{Ballesteros:2012kv}. Then, the EMT \eq{tela} can be written as follows: 
\begin{align} \label{emts}
T_{\mu \nu} = p \, g_{\mu \nu} + \left (\rho + p \right) u_\mu \, u_\nu+\pi_{\mu\nu}\,,
\end{align}
where the non-zero anisotropic stress $\pi_{\mu\nu}$ can be obtained, for instance, from the difference of \eq{emts} and \eq{tela}. The expression \eq{emts} is what in cosmology is usually referred to as an {\it imperfect fluid}. 

The symmetric matrix $B^{ab}$ can be decomposed in terms of its
eigenvectors  $\zeta^a_{(n)}$ and (real) eigenvalues $\lambda_n$,
$n=1,2,3$ as $B^{ab} = \sum_n \lambda_n\, \zeta^a_{(n)} \,
\zeta^b_{(n)}$. Then, we can define the projectors $P^{ab}_{(n)}= \zeta^a_{(n)} \, \zeta^b_{(n)}$ satisfying $\boldsymbol{P}_{(n)} \cdot  \boldsymbol{P}_{(m)} = \delta_{mn} \,\boldsymbol{P}_{(n)}$. The eigenvectors are mutually orthogonal with respect to the three-dimensional metric
$\delta_{ab}$, i.e.\  $\delta_{mn}=\zeta^a_{(m)} \, \zeta^b_{(n)} \, \delta_{ab}$. 
By inspection, $U_{ab}$ is diagonalized by $\zeta^a_{(m)}$ as follows:
\be
U_{ab} =   \delta_{ac} \, \delta_{bd}
\sum_{n=1}^3 \tilde \lambda_n \, {P}_{(n)}^{cd} \,,
 \ee
 where $\tilde\lambda_n= U_{\tau_1} + 2
  \, U_{\tau_2} \, \lambda_n + 3 \, U_{\tau_3} \, \lambda_n^2$.
Therefore, the EMT (\ref{tela}) can be cast in the following form
\be
T_{\mu \nu}= U \, g_{\mu \nu } - 2 \sum_{n=1}^3 \tilde\lambda_n\,\lambda_n\,
\zeta_\mu^{(n)}  \, \zeta_\nu^{(n)} \, , \qquad \zeta_\mu^{(n)} =\lambda_n^{-1/2} \, 
\de_\mu \Phi^a \, \zeta^b_{(n)} \, \delta_{ab} \, .
\ee
Notice that $u^\mu \, \zeta_\mu^{(n)} =0$ and $g^{\mu \nu} \zeta_\mu^{(m)} \,  \zeta_\nu^{(n)}= \delta^{mn}$ and thus, $\{ u_\mu , \, \zeta_\nu^{(m)}  \}$ form an orthonormal tetrad.  The \solid  has  principal pressures given by
\be
p_n \equiv
 T_{\alpha\beta} \, {\zeta^{(n)}}^\alpha \, {\zeta^{(n)}}^\beta=U -2\tilde\lambda_n\,\lambda_n
\, , \quad n=1,2,3 \, .
\ee
The thermodynamics of solids can be studied following analogous lines to those given below and in \cite{Ballesteros:2016kdx} for the special case of perfect fluids. 

\subsubsection{$U(b)$: perfect fluid} \label{Udb}

If the $SO(3)_s$ symmetry of the solid is enlarged to the invariance under volume preserving spatial diffeomorphisms $V_s$Diff, see \eq{mgauge}, the trace operators $\tau_n$ must combine into the determinant $b^2$ defined in \eq{detb}, constraining further the properties of the medium. Indeed, we recall that $b^2$ can be expressed as a function of $\tau_n$ using $b^2=(\tau_1^3-3\,\tau_1\,\tau_2+2\,\tau_3)/6$. Then, under the assumption that only the spatial St\"uckelberg fields are present and imposing $V_s$Diff, the master function $U$ that determines the Lagrangian of the system depends only on  $b$ at LO. However, if the restriction on the field content is relaxed, allowing also for the presence of the temporal St\"uckelberg $\Phi^0$, the operators $Y$ and $X$ should be included as well under at lowest order in derivatives, see Table \ref{tab:summa}. In this section we will focus on the special case $U(b)$ and we will later move progressively to the general structure of the media that respect $V_s$Diff.

In the case $U(b)$ the EMT reads 
\be
T_{\mu \nu}   = p \, g_{\mu \nu} + \left (\rho + p \right) u_\mu \, u_\nu\,, \qquad \rho = -U \,,\quad 
p = U - b \, U_b\,,
\label{EMTpf}
\ee
with $u^\mu$ being defined in \eq{veloc}. In comparison to the general solid \eq{emts}, the symmetry $V_s$Diff removes the anisotropic stress $\pi_{\mu\nu}$, thus restricting the medium to be a perfect fluid. In addition, it also sets to zero the (spin two) graviton mass. 

The application of this type of systems for describing cosmological perfect fluids (and specifically their perturbations) was studied in \cite{Ballesteros:2012kv}. For instance, if the fluid has a barotropic equation of state of the form $p =w \rho$, with constant $w$, the master function is of the form 
\begin{align} \label{backb}
U\propto b^{1+w}\,.
\end{align}
Hence, the {background} dynamics of the $\Lambda$CDM universe can be modelled with
\begin{align}
U\propto(\Omega_r\, (b^{4/3}-1)+\Omega_m\, (b-1)+1)\,,
\end{align}
with $\Omega_r\sim 10^{-4}$ and $\Omega_m\sim 0.26$, to satisfy the current constraints on the radiation and (cold dark) matter densities \cite{PLANCK2}.

In general, the perturbations of a $U(b)$ perfect fluid  can be studied decomposing the spatial phonon fields $\pi^i$, $i=1,2,3$ --introduced in \eq{phonons} for the general four-dimensional case-- into longitudinal and transverse polarizations:
\begin{align}
\pi^i=\pi^i_L+\pi^i_T\,, \quad \partial_i\pi^i_T=0\,,\quad  \epsilon_{ijk}\partial_j\pi^k_L=0\,,
\end{align}
where  the transverse fields $\pi_T^i$ couple to vector metric perturbations and their evolution is dictated by the conservation of vorticity \cite{Ballesteros:2012kv}, which  is a consequence of the symmetry $V_s$Diff. As we discuss in Appendix \ref{conse}, this can be easily obtained applying Noether's theorem, which implies and infinite set of covariantly conserved currents \cite{Dubovsky:2005xd,Ballesteros:2012kv}. 

A peculiar property of the perfect fluids of the kind $U(b)$ is that the transverse modes $\pi^i_T$, which are interpreted as vortices, do not propagate in flat space since the symmetry $V_s$Diff forbids spatial derivatives in their action. Their time evolution is simply given by the equation $\ddot\pi^i_T=0$. This has lead to the suggestion that these fluids may be strongly coupled at all scales \cite{Endlich:2010hf}. However, it has been argued that the issues associated to the evolution of these modes (e.g.\ the appearance of divergences in the scattering of phonons in flat spacetime) can be resolved with a careful choice of the appropriate (physical) observables \cite{Gripaios:2014yha}. 

It can be easily checked, using a field redefinition, that the inclusion of the NLO operators respecting $V_s$Diff and containing only the three spatial St\"uckelberg fields does not modify the evolution equation of $\pi^i_T$ in Minkowski (with respect to the one at LO) \cite{Ballesteros:2012kv}. There are five independent operators at NLO: $\partial_\mu b\,\partial^\mu b$, $H^{\mu\nu}\nabla_\mu u^\alpha\nabla_\nu u_\alpha$, $(u^\mu\partial_\mu b)^2$, $\nabla_\mu u^\nu\nabla_\nu u^\mu$ and $\epsilon_{\alpha\beta\mu\nu}\nabla^\mu u^\alpha\nabla^\nu u^\beta$, see \cite{Ballesteros:2012kv}. Once these operators are  are included, the EMT of the system is no longer that of a perfect fluid. 

The relativistic formulation of self-gravitating media that we are exploring in this paper allows a straightforward connection to the thermodynamic theory of continuous media. This can be done by constructing a dictionary between two EFT pictures: the action \eq{LAll} and the fundamental equations of thermodynamics, see e.g\ \cite{Dubovsky:2011sj}. A more detailed discussion of this dictionary is beyond the scope of this work and and we provide it, for the case of perfect fluids, in \cite{Ballesteros:2016kdx}. For the purpose of illustration we use here the perfect fluid $U(b)$. Essentially, we need to  match the operator $b$
with an intensive thermodynamic variable (or a combination of them). Choosing the particle number density $n$
and the entropy density $s$ as independent variables, we write $b=b(n,\,s)$. Then, the chemical
potential $\mu$ and the temperature are obtained from the first principle of thermodynamics as
$\mu=\partial\rho/\partial n$ and $T=\partial\rho/\partial s$. The pressure enters through the Euler relation
\be
\mu\,s\ +n \, T=\rho+p  \,,
\ee
which must be valid for any $U(b)$. This can be achieved, for instance, with the identification $b=n$ (and then $T=0$), which implies $\mu=-U_b,$ and is in principle 
suitable for a degenerate system (i.e.\ in the limit $T\rightarrow$ 0). A different option is $b=s$ (with
$\mu=0$) and $T=-U_b$, as for a gas of photons. This second choice is the one that was advocated e.g.\ in \cite{Endlich:2010hf,Dubovsky:2011sj,Ballesteros:2012kv}. 

\subsubsection{$U(X)$: perfect fluid} \label{UdX}

At the opposite side of the spectrum from solids and perfect fluids lie the self-gravitating media that can be described at LO by  introducing only the temporal St\"uckelberg field $\Phi^0$.  We have not found a symmetry that forbids the spatial St\"uckelberg fields at LO but reintroduces them at NLO or at higher orders in derivatives. This is analogous to what happens for the solids and perfect fluids studied above, but exchanging the roles of the temporal and spatial fields. By looking at Table \ref{tab:summa}, it seems that $V_s$Diff and $\Phi^a\rightarrow \Phi^a+f^a(\Phi^0)$ select $X$, but this combination of transformations is not a symmetry of an action based on $U(X)$, simply because $\Phi^a$ are not in $X$. Therefore, the absence of these fields must be assumed. 

Clearly, with such an assumption,  the function $U(X)$ is the only shift-symmetric possibility that exists at LO (with a single scalar field). Further restrictions involve necessarily discrete symmetries such as e.g. $X\rightarrow -X$.

Then for a generic $U(X)$, The EMT is given by 
 \be
 T_{\mu\nu}=p \, g_{\mu\nu}+(\rho+p)\vel_\mu\vel_\nu\,,\quad \rho =- U+2 \,X\,U_X \, , \qquad p = U\,.
 \ee
This type of perfect fluid is fundamentally different from the case $U(b)$ because its four-velocity, $\vel^\mu$, is the derivative of a scalar and thus it has zero vorticity. This is due to the fact that for $U(X)$ the transverse phonons $\pi_T^i$ are entirely absent by construction.
Actually, since both $U(X)$ and $U(b)$ describe perfect fluids, their phonon expansions \eq{phonons} can be matched setting to zero the transverse modes of the latter and identifying through an equation the divergence of $\pi^i$ with the time derivative of $\pi^0$. This matching generically involves as well a linear combination of the scalar metric perturbation. The correspondence can also be easily established at the level of the background. For instance, a constant barotropic equation of state is obtained choosing $U \propto X^{(1+w)/(2w)}$, to be compared with \eq{backb}. For an early study of irrotational perfect fluids where longitudinal fluctuations were already interpreted as phonons and a thermodynamic dictionary was provided, see \cite{Matarrese:1984zw}.

In order to extend beyond LO the medium that can be constructed using only the temporal St\"uckelberg, we must linearly add to the action the operators $(\nabla_\mu\mathcal{V}^\mu)^2$ and $\nabla_\mu \mathcal{V}^\nu \nabla^\mu \mathcal{V}_\nu$ at NLO. Clearly, the operator $\nabla^2\Phi^0\sim\nabla_\mu \vel^\mu$ can be omitted assuming appropriate boundary conditions, since it is a total derivative. Besides, the NLO operators $\epsilon_{\mu\nu\alpha\beta}\nabla^\alpha\vel^\mu\nabla^\beta\vel^\nu$ and $\vel^\mu\vel^\nu\nabla_\mu\vel_\alpha\nabla_\nu\vel^\alpha$ need not be included thanks to the fact that $\vel^\mu$ is hypersurface orthogonal. The first of the two vanishes and the second one can be written as a combination of $(\nabla_\mu\mathcal{V}^\mu)^2$ and $\nabla_\mu \mathcal{V}^\nu \nabla^\mu \mathcal{V}_\nu$. 

If we assume that the operator $X$ is absent, and we work only at NLO, the resulting model is a special case of the ``Einstein-aether'' model \cite{Jacobson:2000xp}, see e.g.\ \cite{Blas:2010hb}. 

Notice that the extension to all orders of $U(X)$ is not a generalization of the  ghost condensate \cite{ghost-cond}, but instead the most general modified gravity model based on a single scalar derivatively coupled. Actually, including all the orders in the derivative expansion corresponds to a large class of models contained in the EFT of inflation \cite{Cheung:2007st} (or {DE} \cite{Gubitosi:2012hu}) for which the (soft) breaking of the temporal shift symmetry $\Phi^0\rightarrow \Phi^0+c^0$ (with constant $c^0$) due to a potential term for the inflation (or, generically, the extra scalar mode) is entirely neglected.

\subsection{Four-dimensional media}

We will now move on to describe the media of Table \ref{tab:summa}, which pertain to a different class than the cases we have studied so far, because all of them contain the four St\"uckelberg fields $\Phi^A$, which motivates the name we give them. As we will now see, the most general of these media (whose action is invariant under shifts and internal spatial rotations) is the covariant generalization of a non-relativistic model used in \cite{Son:2005ak} to introduce supersolids. If we further constrain these systems by requiring invariance under internal spatial diffeomorphisms that preserve the volume, we get superfluids \cite{Dubovsky:2011sj}. Symmetries that mix the temporal and spatial St\"uckelberg fields --see Table \eq{tab:summa}-- lead to a set of particular subcases of the general supersolid.

\subsubsection{$U(X \, , Y \,,\tau_n\,,y_n)$: supersolid}

Let us start with the most general media that our symmetries allow. The less symmetric self-gravitating media that we can construct with our assumptions respect only the internal shifts \eq{ssyim} and the spatial rotations $SO(3)_s$ of \eq{rots}. Therefore, their LO master function $U$ contains all the operators of the set \eq{ooops}, that we identified in Section \ref{sym}, and thus their diffeomorphism invariant action is precisely \eq{LAll}. A non-relativistic description of systems characterized precisely by these symmetries was given in \cite{Son:2005ak}, where they were interpreted as zero-temperature supersolids. According to \cite{Son:2005ak}, the three spatial $\Phi^a$ would correspond to the comoving coordinates of the medium (as we have been interpreting them), whereas $\Phi^0$ would be the (shift-symmetric) phase field related to a $U(1)$ symmetry associated to particle number conservation. 

In spite of some claims, see e.g.\ \cite{natsuper}, whether actual supersolids exist in nature is still unclear. For the purpose of this work we will simply use the term to refer to a continuous medium amenable to a coarse grained description with four degrees of freedom and based on the shift and rotational symmetries \eq{ssyim} and \eq{rots}. For a review on supersolids we point the reader to \cite{ssolids}.

The same set of symmetries, equations \eq{ssyim} and \eq{rots}, was later considered in \cite{Nicolis:2013lma}, where the CCWZ method \cite{Callan:1969sn} was applied to derive (for Minkowski spacetime) a relativistic version of the LO action given in \cite{Son:2005ak}. However, the action written in \cite{Nicolis:2013lma} for these symmetries depends on a generic function of all our kinetic blocks $C^{AB}$ except $X$, which is an independent invariant operator under internal translations and $SO(3)_s$ rotations. Besides, the concrete operators $y_n$, defined in \eq{defy}	were not given in \cite{Nicolis:2013lma}.

Phenomenologically, the media of this kind posses many interesting properties. First of all, they share with the solids $U(\tau_n)$ the fact that the graviton acquires a non-vanishing mass that is related to the presence of anisotropic stress and to the speed of propagation of transverse modes. We refer to the group velocity $d\omega/d k$, which given e.g.\ a dispersion relation of the form $\omega^2=k^2+m^2$, clearly involves the mass $m$. In addition, given that $\Phi^0$ and $\Phi^a$ lead to two different four-vectors in the medium, the four-velocities \eq{veloc} and \eq{vdef}, the energy-momentum {tensor} exhibits a non-zero ``heat-flux''.

We point out that the object $C^{AB}$, defined in \eq{Bmatrix}, is the four-dimensional generalization of the left Cauchy-Green tensor, see Section \ref{ssolid}. This agrees with the picture in which the general media described by the EFT \eq{LAll} can be interpreted as four-dimensional media (or hyper-volumes) propagating in spacetime. In Section \ref{massg} we will make the connection between these media and massive (and modified) gravity, using the unitary gauge as tool to simplify the matching.

\subsubsection{$U(X,Y,b)$: superfluid} \label{sflux}

Imposing that the action \eq{LAll} has to be invariant under $V_s$Diff, the internal volume preserving diffeomorphisms  \eq{mgauge}, the only invariant LO operators that appear in the master function are  $X$, $Y$ and $b$. The corresponding EMT is given by
\be
\label{pftd}
T_{\mu\nu}= \left(U - b \, U_b  \right) \,g_{\mu\nu} +\left(Y \,U_Y-b\, U_b
\right) \, u_\mu \, u_\nu + 2 X \, U_X \, \vel_\mu \,  \vel_\nu\,.
\ee
In general this is not a perfect fluid, due to the four-velocities $\vel^\mu$ and $u^\mu$ not being parallel. The fact that $\vel^\mu$ has zero vorticity has motivated interpreting these media as relativistic superfluids \cite{Dubovsky:2011sj, Nicolis:2011cs}, following earlier ideas for (non-relativistic) supersolids \cite{Son:2005ak} and superfluids \cite{Son:2002zn}. The key idea of this picture is that the EMT  \eq{pftd} can be seen as a two-component fluid whose admixture allows to describe the various phases of a superfluid. For instance,  it is straightforward to show that the phonon expansion of the action leads to two kinds of longitudinal waves (propagating with different speeds of sound), which suggests interpreting them as the first and second sound in superfluids \cite{Nicolis:2011cs}.
\newline

\noindent It is worth highlighting some special cases of the general superfluid: 
\begin{itemize}
\item If in addition to the symmetry $V_s$Diff we impose also invariance under $\Phi^0\rightarrow \Phi^0+f(\Phi^0)$, the master function has to be a function of the operators $\mathcal{O}_\alpha$, which we defined in \eq{On}, and $b$. The operators $\mathcal{O}_{-1}$ and $b$, together with the NLO operators that can be constructed from $\mathcal{V}^\mu$, were used in \cite{Blas:2012vn} to model Lorentz breaking in dark matter. 
\item The master function {$U(b,Y)$} describes a perfect fluid with four-velocity $u^\mu$ and energy density and pressure 
\begin{align} \label{nbar}
\rho=Y \, U_Y-U\,,\quad p= U-b\, U_b\,.
\end{align} 
This case is interesting because it comes from imposing not only the symmetry $V_s$Diff but also requiring invariance under $\Phi^0\to \Phi^0+f(\Phi^a)$, see Table \ref{tab:summa}. This feature makes it qualitatively different from the other perfect fluids that we have encountered: $U(X)$ and $U(b)$, whose LO structure does not derive from symmetries acting on the four St\"uckelberg fields. 
One can obtain a constant barotropic equation of state for $U(Y,b)$ choosing 
\begin{align} \label{exp}
U\propto b^{1+w}\;{\mathcal U}(\,b^{-w}\,Y)\,,
\end{align}
where $\mathcal U$ is an arbitrary function. Another choice of master function which also leads to a constant barotropic equation of state is $U\propto (b^{1+w}+Y^{1+1/w})$.
\item An equation of state $w=-1$ can be obtained from \eq{exp} by choosing 
\begin{align}
\mathcal{U}=\mathcal{U}(b\, Y),
\end{align} 
which interestingly corresponds to the enhanced symmetry
\begin{align}
\Phi^A\to\Psi^A(\Phi^B)\,,\quad \det |\partial \Psi^A /\partial{\Phi^B}|=1\,.
\end{align}
By looking at equation \eq{vicev}, it is now clear why $b\,Y$ is the operator that allows to switch from Eulerian to Lagrangian coordinates.

\end{itemize}

\subsubsection{Special supersolids} \label{sss}

These media are the subclasses of the general action \eq{LAll} listed in Table \ref{tab:summa}. 

\begin{itemize}
\item For instance, imposing the symmetry $\Phi^0\to \Phi^0 +f(\Phi^a) $, the leading invariant operators are
$\tau_n$  and $Y$ and the EMT is
\be
T_{\mu\nu}=U \,  g_{\mu \nu}  -2 \, U_{ab} \, \de_\mu \Phi^a \de_\nu
\Phi^b  + Y \, U_Y \, u_\mu \, u_\nu \, . 
\label{scf}
\ee
Given what we have seen so far, a natural interpretation of \eq{scf} seems to be that it describes a solid coupled to a perfect fluid. An analysis of the thermodynamics of perfect fluids \cite{Ballesteros:2016kdx} suggests that $Y$ can be interpreted as the temperature so that $U(\tau_n, Y)$ may be a possible description for a solid at finite temperature.
 
\item The symmetry $\Phi^a \to \Phi^a + f^a(\Phi^0)$ enforces the master function to be of the form $U= U(w_n,\,X)$. If in addition we impose the symmetry $\Phi^0\rightarrow\Phi^0+f(\Phi^0)$, the operators $w_n$ are selected. We will comment on these cases in the next section, where we build the connection between massive gravity and self-gravitating media. 

\end{itemize}

\section{SO(3) spatially symmetric massive gravity} \label{massg}

The main motivation for massive and modified gravity in the
context of cosmology is the possibility that the observed acceleration
of the Universe could be  due to a modification of GR that weakens gravity at large
distances. In this respect, the idea of massive gravity (MG) is appealing, since endowing the
graviton with a mass could effectively make gravity a short range
interaction (on cosmological scales).

The first attempt to provide a mass to the graviton dates back to 1939; when Fierz and Pauli  \cite{Fierz:1939ix} studied the most general Lorentz invariant mass term for metric perturbations around Minkowski space.  
It was thought for a long time that any extension of the Fierz-Pauli Lagrangian beyond the the quadratic level was pathological. Indeed, Boulware and Deser argued that at the non-linear level a ghost mode would have necessarily be present \cite{BD}. The problem was solved in \cite{Gabadadze:2011} by finding a ghost free extension of the Fierz-Pauli Lagrangian
at the non-linear level \cite{Gabadadze:2011, HR}. This model, which exhibits Lorentz symmetry when expanded around Minkowski, solves the ghost issue at the (heavy) price of lacking spatially flat homogenous FLRW solutions \cite{DAmico}. As shown in \cite{ArkaniHamed:2002sp} the Fierz-Pauli model as an effective theory (and also \cite{Gabadadze:2011}) has a (very low) cut-off $\Lambda_3 = (m^2\,M_{pl})^{1/3}$, where $m$ sets the scale of the graviton mass mass scale.  A recent proposal to improve this cut-off has been given in \cite{deRham:2015ijs,deRham:2016plk}. 

Fierz-Pauli MG fails to reproduce the correct light bending around heavy and dense objects, in sharp contrast with standard GR.  This also happens in the limit of vanishing graviton mass, an issue that is known as the vDVZ discontinuity \cite{DIS}.  A possible way out was proposed by Vainshtein
in \cite{Vainshtein}, by arguing that  non-linear effects restore the correct GR behaviour at short distances from a gravitational source.  
Whereas this mechanism has been studied for various models \cite{vain2}, it is important to stress that it relies on strong non-linearities; even at the macroscopic scales of the Solar System where the value of the gravitational potential is small. 

A possible way out of this last difficulty is to abandon Lorentz invariance in the gravitational sector\footnote{Bounds on Lorentz violation in the gravitational sector are rather mild assuming the equivalence principle and come from post-Newtonian preferred frame effects, see e.g.\ \cite{Blas:2014aca}, and gravitational waves emission.} in favour of a simpler symmetry group; in practice, by requiring only spatial rotational invariance~\cite{Rubakov,Dubovsky:2004sg}.  All Lorentz-breaking MG models with five DoF having (at least) spatial rotational invariance are free of ghost instabilities and can be classified \cite{canonical-us,general-us}. In addition, they have no vDVZ discontinuity and FLRW solutions do exist for these models \cite{cosmo5dof}.  Their range of validity is given by an ultraviolet
cut-off $\Lambda$, which is dictated by the 
symmetries that define the EFT. Typically, for Lorentz-breaking MG  the cut-off is $\Lambda_2= (m \,
M_{pl})^{1/2}\gg \Lambda_3$ \cite{Rubakov,Rubakov:2008nh,Dubovsky:2004sg,cosmo5dof}. In order to have any impact on the current acceleration of the Universe, the scale $m$ should thus be of the order of today's Hubble scale, giving  $\Lambda_2^{-1}\sim 0.1$mm.

The EFT of four derivatively coupled scalar fields with internal spatial $SO(3)_s$ invariance allows to re-interpret Lorentz-breaking MG models as self-gravitating media. From the perspective of this EFT, the breaking of Lorentz symmetry is simply a consequence of a background choice. MG is then a class of models within a broad framework. In other words, the EFT of self-gravitating media leads to a systematic construction of models for the acceleration of the Universe, including MG. 

The basic idea that helps to build this connection is the St\"uckelberg {\it mechanism} or {\it trick}, which we already mentioned in Section \ref{Aseqm}. The application of the St\"uckelberg mechanism for Fierz-Pauli MG
was already given in \cite{Siegel:1993sk} and was later generalized
to other models of MG in \cite{ArkaniHamed:2002sp} and
\cite{Dubovsky:2004sg}. If $h_{\mu\nu}$ is a metric perturbation
around a (reference) background metric, $\bar g_{\mu\nu}$, such that
the full metric is $g_{\mu\nu}=\bar g_{\mu\nu}+h_{\mu\nu}$, a generic
(Lorentz invariant) MG potential may be expressed as
function of the perturbation $h_{\mu\nu}$ with indices raised using
$g^{\mu\nu}$, i.e.\ $\sqrt{-g}\,V(h_{\mu\nu}, g^{\mu\nu})$. Given such
a potential, added to the Einstein-Hilbert Lagrangian, the
St\"uckelberg trick can be implemented replacing the background metric
$\bar g_{\mu\nu}$ with a tensor field made out of four scalars:
$\mathcal{G}_{\mu\nu}\equiv\partial_\mu\Phi^A\partial_\nu\Phi^B\bar
g_{AB}[\Phi^C(x^\alpha)]$, so that the metric perturbation
$h_{\mu\nu}$ is replaced with $g_{\mu\nu}-\mathcal{G}_{\mu\nu}$, see
e.g.\ \cite{Hinterbichler:2011tt}. Using these replacements to express
the action in terms of the spacetime metric $g_{\mu\nu}$ and the
tensor $\mathcal{G}_{\mu\nu}$ we obtain a covariant embedding of any
Lorentz invariant MG model originally defined with a
reference metric $\bar g_{\mu\nu}$. This procedure can be adapted to
deal also with Lorentz breaking models, as described in
\cite{Dubovsky:2004sg}. 

\begin{table}[t]
  \footnotesize
  \small
  \renewcommand\arraystretch{1.5}
  \centering
  \begin{tabular}{|c|c|c|}
    \hline
{\bf LO self-gravitating media} & Map   &{\bf Massive gravity}  \\ 
 ${\cal L}(C^{AB},\,g_{\m\n})$&  Unitary gauge {\color{red}$\;\; \longrightarrow$} & ${\cal L}(h_{\mu\nu},\, g^{\mu\nu})$          \\ 
 $\mathcal{O}_{LO}:\,X,\,Y,\,\tau_n,\,y_n$&{\color{red}$\longleftarrow$}\,\, St\"uckelberg ``trick"&  $SO(3)$ invariants of ADM's $N,\,N^i,\,\gamma_{ij} $      \\ \hline
       \end{tabular}
  \caption{\it Relationship  between material Lagrangians and massive gravity models.}
  \label{tab:sumx}
\end{table}

In this work, we started our analysis directly
with an explicitly diffeomorphism invariant action \eq{saction} in a ``top-down'' approach \cite{Rubakov:2008nh}. This
automatically implements the St\"uckelberg mechanism for the
actions obtained imposing the background $\Phi^A = \delta_\mu^A
x^\mu$. Indeed, assuming that the condition $\det(\partial_\mu \Phi^A)\neq 0$ holds, it is possible to find a local set of coordinates (unitary gauge) such that
\be
\de_\mu \Phi^A = \delta_\mu^A
\, .
\label{ugauge}
\ee
Notice that, defined in this way, the unitary gauge is actually a collection of gauges, which are equivalent modulo constant shifts, i.e.\ $\Phi^A=x^A+c^A$ with $\partial c^A/\partial x^\mu =0$, where for simplicity we do not distinguish between spacetime and internal manifold indices. These gauges are all equivalent since the action is, by assumption, shift invariant in the St\"uckelberg fields.

The unitary gauge reflects the fact that since we have four scalars $\Phi^A$ and four spacetime coordinates, we can use the diffeomorphism invariance of the action \eq{LAll} to absorb the dynamics of the scalar fields into the degrees of freedom of the metric. 
By using the ADM splitting of
spacetime to describe the metric $g_{\mu\nu}$ and its
perturbations \cite{Arnowitt:1962hi}, it is then straightforward to  make a connection with MG in the so-called broken
phase. Therefore, this phase (of broken diffeomorphisms) can be simply understood as a way of expressing the dynamics of the comoving coordinates of the self-gravitating medium through their effect on the metric perturbations. The correspondence between self-gravitating media and massive and modified gravity models written explicitly in terms of metric perturbations is depicted in Table \ref{tab:sumx}, where the arrows indicate how to move from one picture to the other. 

In the ADM formalism the metric can be written in terms of the lapse $N$, the
shifts $N^i$ and the spatial metric $\gamma_{ij}$, namely 
\be
g_{\mu \nu} = \begin{pmatrix} -N^2 + N^i \, N^j \, \gamma_{ij} &
  \gamma_{ij} \, N^j \\
 \gamma_{ij}  \, N^j & \gamma_{ij} \end{pmatrix} \, , \qquad 
 g^{\mu \nu} = \begin{pmatrix} -1/N^2 &   \, N^i/N^2 \\
   \, N^j /N^2 & \gamma^{ij} - N^i \, N^j /N^2 \,
    \end{pmatrix}\,,
\ee
where $\gamma^{i m} \gamma_{m j} = \delta^i_j$. In the unitary gauge \eq{ugauge},  the matrix $C^{AB}$ defined in (\ref{Bmatrix}) coincides with inverse of
the spacetime metric:
\begin{align}
C^{AB}=g^{\mu\nu}\delta^{A}_\mu\delta^B_{\nu}\,.
\end{align}
Defining 
\begin{align}
\xi^i\equiv {N^i}/{N}\,,
\end{align}  
for convenience, we decompose $C^{AB}$ in the unitary gauge as follows:
\be \label{asfll}
X=C^{00} = - 1/N^{2} \, , \quad \mathcal{C}^i= C^{i0} =  \xi^i/N \,,\quad
B^{ij}=C^{ij}= \gamma^{ij}- \xi^i \, \xi^j \,.
\ee
Then, we can write the building blocks of LO self gravitating media --see \eq{ooops} and Table \ref{tab:operators}-- in terms of the ADM variables:
\begin{align}
X=-1/N^2\,,\quad Y=\frac{1}{N\,\sqrt{1-\xi^2}}\,,\quad \tau_n=\text{Tr}\left[\left(\boldsymbol{\gamma}-\xi\otimes\xi\right)^n\right]\,,\quad  
 y_n=\frac{\xi\cdot(\boldsymbol{\gamma}-\xi\otimes\xi)^n\cdot\xi}{N^{2}}\,,
\end{align}
where $\boldsymbol{\gamma}$ is the $3\times 3$ matrix of components $\gamma^{ij}$. In these expressions the dot ($\cdot$) represents the standard matrix product and we use the following notation: $\xi^2\equiv\xi^i\,\gamma_{ij}\, \xi^j$ and $(\xi\otimes\xi)^{ij}\equiv\xi^i\xi^j$. Therefore, the master function $U$ of the action \eq{LAll} becomes  an algebraic function of the ADM variables in the unitary gauge:
\be
S=  M_{pl}^2 \int d^4x \, \sqrt{-g}\, R + \int d^4x \, \sqrt{-g}\,U \left(N,\xi^i, \gamma^{ij} \right) \,.
\label{ugaction}
\ee
Special cases --see Tables \ref{tab:summa} and \ref{3dimmedia}-- can be obtained from the remaining $SO(3)_s$ invariant scalar operators of Table \ref{tab:operators}, which are combinations of $X$, $Y$, $\tau_n$ and $y_n$:
\begin{align}
\begin{aligned}
b=\sqrt{(1-\xi^2)\det\boldsymbol{\gamma}}\,,\quad  w_n={\text{Tr}\left(\boldsymbol{\gamma}^n\right)}\, ,\quad
 {\cal O}_{\alpha\beta
   n}=(-1)^\alpha \left(1-\xi^2\right)^{\alpha+\beta}\left(\xi\cdot(\boldsymbol{\gamma}-\xi\otimes\xi)^n\cdot\xi\right)^\beta\,.
\end{aligned}
\end{align}

Notice also that the four-velocities $\vel^\mu$ and $u^\mu$ become in the unitary gauge:
\be
u^{\mu}=\frac{\delta^\mu_0}{N\,\sqrt{1-\xi^2}}\,,\qquad \vel^{\mu}=(-1/N,\,\xi^i)\,.
\ee
If the the shifts are zero, e.g.\ as in  Minkowski and  FLRW   spacetimes, $u^\mu$ and $\vel^\mu$ coincide, up to a sign; see also Section \ref{Media}.

The action \eq{ugaction} was the starting point in \cite{canonical-us,general-us,nonder-us} where a large class of non-derivative\footnote{By ``non-derivative" we refer to the actions for which the master function $U$ contains only non-derivative combinations of the ADM variables, precisely as in \eq{ugaction}.} massive and modified gravity models --those with spatial $SO(3)$ invariance-- was studied by means of a Hamiltonian analysis. Models with 6, 5, 3 and 2 degrees of freedom (DoF) were thus found. The condition of $SO(3)$ spatial invariance turns to be important to avoid ghost instabilities. If this restriction is relaxed, 6 DoF generically propagate and one of them (a scalar mode with respect to spatial rotations) is a ghost around flat space, which corresponds to the infamous Boulware-Deser ghost \cite{BD}. 
 
For example, a $SO(3)$ invariant master function $U$ of the form
\be
U =   \,\U(\tau_n ) +\sqrt{-X}\;\mathcal{E}(w_{n})
\label{5dof}
\ee
propagates precisely 5 DoF \cite{general-us} (provided that $\mathcal U$ is not a constant). As a matter of fact, this not the most general $U$ that propagates 5 DoF, see \cite{general-us}. We leave open a detailed exploration of the structure of the most general case, which is anyway contained in our formalism, for a possible future investigation. The functions $\U$ and $\mathcal{E}$ of \eq{5dof} are generic and thus there exist
infinitely many $U$ with  5 DoF. Therefore, according to the discussion in the previous section ghost-free massive gravity can be identified with a specific kind of supersolid.
Moreover, setting $\mathcal{E}=0$ there are still 5 DoF and the medium is a solid. Thus, the construction of ghost-free massive gravity with five DoF needs at least the three spatial St\"uckelberg fields $\Phi^a$. 

Cases where only 3 or 2 DoF occur also exist \cite{nonder-us}. In particular, exactly 3 DoF are present if $U$ has the  structure
\be
U= U(w_n,\,X)\,.
\label{3dof-1}
\ee
We recall that the above form for $U$ is protected by the symmetry: $\Phi^a \to \Phi^a + f^a(\Phi^0)$, see Table
\ref{tab:summa}. It should be stressed that although 3 DoF are present at the non-perturbative level for \eq{3dof-1}, only two  DoF are found expanding around flat space \cite{nonder-us}.  A proposal for a UV completion,  where the LO order part of the action is built out of the scalar operators $w_1$ and $w_2$, was put forward in \cite{Blas:2014ira} adding the symmetry $\Phi^0\rightarrow \Phi^0+f(\Phi^0)$, see Table \ref{tab:summa}. In that model   the fields $\Phi^a$ are further coupled with a triplet of purely spatial  vector fields, whereas the dynamics $\Phi^0$ is dictated by NLO operators.

A particular case of \eq{3dof-1} is given by the Lagrangian
\be
U=\sqrt{-X}\,\mathcal{E}(w_n)+\lambda\,,
\label{3dof-2}
\ee
where only 2 non-perturbative DoF are found. In this expression $\lambda$ is a (dimensionless) cosmological constant. This last case is rather interesting since it gives an example of a
model of gravity with 2 DoF that is different from GR. 

\section{Cosmological perturbations in FLRW} \label{cosmox}
In this section we discuss a basic aspect of cosmological perturbation theory for self-gravitating media. In particular, we are interested in determining the number of degrees of freedom that propagate in a FLRW background for the different kinds of self-gravitating media that we have considered. A more detailed
account of the cosmology of these media will be given elsewhere~\cite{future-cosm}. Here we present the Lagrangians for perturbations that are required for such a study and discuss their main features.

Let us consider a spatially flat\footnote{Spatial curvature  can be introduced without altering the conclusions. For simplicity we set it to zero.} FLRW metric written in the following way:
\be
ds^2 = - N(t)^2\, dt^2 + a(t)^2 \, \delta_{ij} dx^i dx^j \,,
\label{FRW}
\ee
where we denote $x^0=t$. The derivatives with respect to $t$ will be indicated by primes. For instance, by definition $\mathcal{H}=a'/a$ is the Hubble function in these coordinates. With this choice of metric, the following field configuration: 
\be \label{cbasis}
\Phi^0 =\phi(t) \,, \quad \Phi^i = x^i
\ee
is compatible with the equations of motion for the St\"uckelberg fields derived from the action \eq{LAll}, giving a spatially homogeneous and isotropic EMT. In what follows we will work in the unitary gauge, setting $\phi(t) = t$. It is important to stress that in general it is not possible to choose $N(t)=a(t)$ or $N(t)=1$ at the same time that $\phi(t) = t$. Therefore, $N(t)$ has to be determined using the (background) equations of motion. 

{We will now study perturbations around the metric \eq{FRW}. Choosing the unitary gauge, the dynamics of perturbations for the whole system (metric and St\"uckelberg fields) is encoded in the metric.\footnote{In the unitary gauge, the ``phonons" $\pi^A$ of \eq{phonons} are set to zero. We could as well choose to trade some degrees of freedom of the metric \eq{spatg} by the $\pi^A$.} Denoting the metric perturbations by $h_{\mu\nu}$, the master function $U$ of the action \eq{LAll} can be expanded up to second order in $h_{\mu\nu}$ as follows:}
\be \label{masst}
\begin{split}
&\sqrt{- g} \, U=  t^{\mu \nu} \, 
  h_{\mu \nu} 
  + \frac{M_{pl}^2} {4} \left(m_0^2 \,  h_{00}^2+2\,m_1^2\,  h_{0i}
    h_{0i}   -2\,m_4^2 \,  h_{00}\,  h_{ii} +m_3^2 
\,  h_{ii}^2  - m_2^2\,   h_{ij} \,  h_{ij}\right)  \,,
\end{split}
\ee
where $t^{\mu\nu}$ and the masses $m_i^2$ depend on $U$ and its derivatives.

Although the operator $b$ is redundant once the $\tau_n$ are included, it is convenient to single out its effect, since it plays a special role for some symmetry choices. Hence, we write the LO action \eq{LAll} as
\begin{align} \label{LAllb}
S=M_{pl}^2 \, \int d^4 x\,\sqrt{-g}\, R+\int d^4x\sqrt{-g} \, U(X, \, Y, \,  \tau_n, \, y_n,\,b)\,.
\end{align}
Then the masses of \eq{masst} are given by:
\begin{align}
 m_0^2 & =  -\frac{a^7}{ 2 \, N^3 \, M_{pl}^2}  \left[\frac{4 \, 
   U_{XY}}{N^{3}}-\frac{4 \, U_{XX}}{N^{4}}- \frac{{U_{YY}-4 \,
     U_X}}{N^{2}}- 
\frac{ U_Y}{N}+U \right]\, , \\
 m_1^2 & =  \frac{a}{N \,  M_{pl}^2 }\left[a^4 \, U +\frac{2 \, a^4}{N^2}  U_X-a \, U_b+ \sum_{n=0}^3  2 n \, a^{2 -2 n}  \,
   U_{y_n} -2 
   \sum_{n=1}^3  a^{4-2\,n} \, U_{\tau_n} \right] ,\\
 m_2^2 & =  \frac{N}{  M_{pl}^2 }\left[ a^3 \, U -  U_b  - 4
  \sum_{n=1}^3 (n+1) \,  \, a^{3 - 2\,n} \, U_{\tau_n}
\right ] \, ,\\ \nonumber 
 m_3^2 & =   \frac{N}{ M_{pl}^2 }\Bigg[ \ha \left( a^3 \, U +a^{-3} \,
    U_{bb} - U_b \right) +8 \, a^{-3} \, U_{\tau_1 \tau_2} 
  +12 \, a^{-5} \, U_{\tau_1 \tau_3}+
24 \, a^{-7} \, U_{\tau_2 \tau_3}\\
& +\left. 2 \,  \sum_{n=1}^3 \left( n \, a^{-2\,n} \, U_{b \tau_n} -  n \, a^{3- 2\,n} \,
U_{\tau_n}  + n^2
  \,  a^{3-4\,n} \, U_{\tau_n \tau_n} \right) \right]  \, ,\\ \nonumber
 m_4^2 & =  \frac{a^2}{2 \, N^3  \,  M_{pl}^2 }\Bigg[{a^3 \, N^2 \,
  U-a^3 \, N  \, U_Y+2 \, a^3 \, U_X- N^2
 \,   U_b - 2 \,  U_{bX}+ N \, U_{bY}} \\
&  + 2  \sum_{n=1}^3   n \,  \, a^{3- 2\,n} \left( N \, U_{Y \tau_n} -  \,N^2 \, U_{\tau_n} -2 \,  U_{X \tau_n}\right)  \Bigg]\, .
\end{align}

These masses can be simplified (on a case by case basis) taking into account the background equations of motion. Concretely, the condition $t^{\mu\nu}=0$ is required for {consistency. This} gives two equations:
\begin{align}\label{onebkga}
\begin{aligned}
6\, M_{pl}^2 \,\mathcal{H}^2+N^2 \,U-N \,U_Y+2 \,U_X & = 0\,,  \\
N^3U+2 \, M_{pl}^2 \, N \, \left( 
   2 \, \mathcal{H}'+3 \, \mathcal{H}^2\right) -4 \,M_{pl}^2  \,{\cal H}
 \, N' 
-2 N^3 \, \sum_{n=1}^3 n \, a^{
   -2\,n}\, U_{\tau_n}-a^{-3} \, N^3 \, U_b & = 0\,.
   \end{aligned}
\end{align}
These equations are equivalent to the conservation of the medium's EMT on the background or, equivalently, to the equations of motion of the
St\"uckelberg fields (in the background), see Section \ref{Aseqm}. Eliminating $\mathcal{H}'$ from the second of them we obtain the condition
\begin{align}\label{bianchi}
F_1(U,N,a) \, N' + {\cal H} \, F_2\,(U,N,a) =0\,,
\end{align}
where
\begin{align}
F_1 & = - \frac{a^3}{2 \, N^4 \, {\cal H}} \left( N^2 \, ( U_{YY} - 2\, U_X) -
  4 \, N \, U_{XY}+ 4 \, U_{XX}  \right) \,,\\
  F_2 & = \frac{3}{2}\,\left(a^3 \, U_{Y} - U_{bY}  -2\, \sum_{n=1}^3\, n \, a^{3-2\,n} \,
  U_{\tau_n} \right) %\\& 
  +\frac{3}{N}\, \left( U_{bX} - a \, U_X +2 \,\sum_{n=1}^3\, n \, a^{3-2\,n} \,
  U_{X \,\tau_n}\right)\,.
\end{align}
Therefore,  we have the following possibilities:
\begin{itemize}
\item $F_1 \neq 0$ and therefore it is possible to solve for $N'$. As a result, $N$ is dynamical.
\item  $F_1 $ is identically zero. Therefore, (\ref{bianchi}) is an
  algebraic equation for $N$, i.e.\ $F_2(U,\,N,\,a)=0$, except if ${\cal
    H}=0$. 
\item  Both $F_1$ and $ F_2$ are identically zero. In this case $N$ is not fixed by the background (so there exists a residual gauge freedom). We can fix $N$, for instance, to be equal to $a$. 
\end{itemize}

{Let us now consider the scalar, vector and tensor modes of the metric with respect to spatial rotations. A general perturbation of \eq{FRW} can be written as:
\be
ds^2 = - N(t)^2 dt^2 + a(t)^2 \, \delta_{ij} dx^i dx^j + a(t)^2 h_{\mu \nu}(x^\alpha) \, dx^\mu dx^\nu \, , 
\ee
where
\be \label{spatg}
h_{00} = \psi \, , \qquad h_{ij}= \chi_{ij} + \de_i s_j + \de_j s_i + \delta_{ij} \, \tau +
\de_i \de_j \sigma , \qquad h_{0i} = u_i+
\de_i v  \, ;
\ee
with
\be
\de_i u_i = \de_j \chi_{ij}= \de_i s_i=0 \, , \qquad \delta_{ij}\chi_{ij}=0 \,.
\ee
In the scalar sector,} one can integrate out the fields {$v$} and $\psi$ of \eq{spatg}, arriving to  the
following canonical form for the effective quadratic Lagrangian
\be
 L_{(2)}^{(s)}= \frac{1}{2} {\varphi'}^t \, K \, \varphi' +  {\varphi'}^t \, D \varphi -
\frac{1}{2}  {\varphi}^t \, A_{\text{mass}} \,\varphi \, , \qquad
\varphi^t = (\tau \, , - \Delta \sigma ) \, ,
\ee
where the $2\times 2$ matrices $K$, $D$ and $A_{\text{mass}}$ satisfy  $D^t=-D$, $K^t=K$ and $A_{\text{mass}}^t=A_{\text{mass}}$, respectively. 
The propagation of modes is related to the determinant of the matrix $K$:  if $\text{Det}(K) =0$ there can be at most one propagating scalar DoF.
We have that 
\be
\det(K) \propto \left(6 \,a^5 \,\mathcal{H}^2+m_1^2 \,N^3\right) \left(3\, a^7\,
   \mathcal{H}^2-m_0^2 \, N^5\right)
\, .
\label{det}
\ee
Defining the effective masses
\be
M_{0}^2 = m_0^2 \, N^5 -3\, a^7
   \mathcal{H}^2\,,\quad
M_{1}^2 = 6 \,a^5 \,\mathcal{H}^2+m_1^2\, N^3\,,
\ee
we see that if $M_{0}^2=0$ or $M_{1}^2
= 0$, there is a single propagating scalar mode; whereas if $M_{1}^2= M_{0}^2=0$ no scalar
mode  propagates. In a general case, two different scalar modes can propagate. The study of scalar perturbations is important for cosmological signatures that could help to distinguish a modification of gravity such as the ones discussed here from a pure cosmological constant, see \cite{Amendola:2016saw} for the prospects with the Euclid satellite. First, given a certain background evolution, the growth of dark matter perturbations is generically altered if there are extra scalar modes. This can be detected by measuring precisely  the so-called growth index and its scale and redshift dependence, see e.g. \cite{  Wang:1998gt,Linder:2005in,Bertschinger:2008zb,Ballesteros:2008qk,Perivolaropoulos:2008ud}. Moreover, a detection of the speed of propagation of {DE} perturbations would also indicate a deviation from the $\Lambda$CDM model, see e.g. \cite{  Bean:2003fb,Hannestad:2005ak,Takada:2006xs,dePutter:2010vy,Ballesteros:2010ks,Sapone:2010iz}. Besides, a late-time measurement of a non-negligible excess in the gravitational slip (i.e.\ the difference between the two, gauge invariant, Bardeen potentials \cite{Bardeen:1980kt}) would signal the presence of an anisotropic stress component, see e.g.\ \cite{Amendola:2007rr,Bertschinger:2008zb,Daniel:2008et}. In our case, a gravitational slip appears in solids and supersolids, for which (at linear order in perturbations)
\begin{align} \label{anis}
\delta T_{ij}\supset M_2^2\, \partial_i\partial_j \pi_L\,,
\end{align}
where $\pi_L$ is defined via $\pi^i_L=\partial_i\pi_L$ and $\pi^i=\pi^i_L+\pi^i_T$, being $\epsilon_{ijk}\partial_j \pi^k_L=0$, {see \eq{phonons},} and we define:
\be
M_{2}^2\,M_{pl}^{2} =  \sum_{n=1}^3 n^2
  \, a^{-2(n-1)} \, U_{\tau_n} \, .
\ee
Notice also that in a general case, the two scalar DoF that appear from the action \eq{LAllb} allow for the propagation of an adiabatic mode and a entropy mode, see also \cite{Ballesteros:2016kdx}.

\begin{table}[t]
\footnotesize
  \small
  \renewcommand\arraystretch{1.5}
  \centering
\begin{tabular}{|c|c|c|c|c|c|c|} 
\hline
{\bf Media} &{\bf Operators}& $\text{Det}(K)$ &  $M_{0}^2$&
$M_{1}^2$& $M_{2}^2$ & DoF \\ \hline \hline
\multirow{4}{*}{perfect fluids} & $ b$ & $0$  & $0$ & $ \neq 0$ &  $0
                                                                  $  &
                                                                       3    \\ \cline{2-7} 
& $ X$ & $ 0$   & $\neq 0$ &  $0$& $0$ &  3    \\ \cline{2-7}
 & $ Y$ & $\neq 0$ &$\neq 0$ & $\neq 0$  & 0 &    3  \\ \cline{2-7}  
 & $ b,\;Y$ & $\neq 0$ &$\neq 0$ & $\neq 0$  & 0 &   3    \\ \hline  
\multirow{2}{*}{superfluids}& $ b,\;X$ & $\neq 0$ &$\neq 0$ & $\neq 0$  & 0 &   4   \\ \cline{2-7}  
 & $ b,\;X,\;Y$ & $\neq 0$ &$\neq 0$ & $\neq 0$  & 0 &  4    \\ \hline  
solid  & $\tau_n$  & $ 0$ &$ 0$ & $\neq 0$  & $\neq 0$ & 5    \\\hline  
\footnotesize{special supersolid} & $\tau_n,\;Y$  & $ \neq 0$ &$ \neq 0$ & $\neq 0$  & $\neq 0$ & 6    \\ \hline
\footnotesize{special supersolid} & $w_n,\;X$  & $  0$ &$ \neq 0$ & $ 0$  & $\neq 0$ & 2    \\ \hline    
\end{tabular}
\caption{\it Counting of degrees of freedom (DoF) for various LO self-gravitating media in FLRW.}
\label{T3}
  \end{table}

In the tensor sector there are always two propagating DoF, with quadratic Lagrangian
\be \label{GW}
L_{(2)}^{(t)}= \frac{M_{pl}^2}{2} \left[ \frac{a^3}{ N}
  \, \chi_{ij}'
\chi_{ij}' +N \, a 
\left( 2 \, M_{2}^2  - k^2 \right) \chi_{ij}  \,
\chi_{ij}\right ]  \,.
\ee
A non-zero $M_2^2$ changes  the propagation speed of gravitational waves with respect to that of light. This could induce, for instance, observable effects on the propagation and lensing of CMB $B$-modes if the continuous medium is relevant at sufficiently early times, see \cite{Amendola:2014wma,Raveri:2014eea}. It is remarkable that the same quantity, $M_2^2$, that is responsible for the gravitational slip also determines the propagation of gravitational waves. The relation between these two effects has also been noted for $V$Diff systems of three scalars at NLO in derivatives \cite{Ballesteros:2014sxa} and for other examples of modified gravity models \cite{Saltas:2014dha}.
 
Finally,
in the vector sector the quadratic Lagrangian reads
\be
L_{(2)}^{(v)}=  M^2_{pl} \left[\frac{a^3 \, k^2}{2 \, N} (u_i -s_i')  (u_i -s_i') +
\frac{ M_{1}^2}{2 \, N^3} \, \, 
u_i u_i + a \, N \, k^2   \, M_{2}^2  \, s_i \,  s_i \, \right].
\ee
The fields $u_i$ have a purely algebraic equation of motion and thus can be integrated out, 
giving the Lagrangian
\be
\tilde{L}_{(2)}^{(v)}=k^2 \,  M^2_{pl} \left [\frac{a^3 \,  M_{1}^2 }{{2\,
   a^3 \,N^3 \, k^2+2\, N \,M_{1}^2}}\;s_i's_i'+ N \, a  \,
M_{2}^2  \, s_i \,  s_i \, \right] \,. 
\ee
 The vectors $s_i$ propagate only if
$M_{1}^2 \neq 0$, see Table \ref{T3}. The dispersion relation is not
trivial only when $M_{2}^2\neq0$. Thus,  $M_{2}^2
$, besides controlling the dispersion relation of tensors, also
determines the dynamics of the vectors.
The fact that   $M_{2}^2 \neq 0$  in the unitary gauge is
equivalent to the presence of an anisotropic stress in the EMT of the 
St\"uckelberg fields.
As we mentioned above, a complete study of the cosmology of these media will be given in a separate
publication \cite{future-cosm}. We summarize in
Table \ref{T3} the DoF counting for various examples of sets of operators that may appear inside the LO master function.

\section{Dark matter and dark energy signatures} \label{sigs}

To illustrate the novel phenomenology that arises for the description of dark energy and dark matter from the effective theory of self-gravitating media, we will focus now in certain aspects of cosmological perturbations that are specific of our framework. In particular, we will consider scalar perturbations for non-barotropic fluids and, also, the propagation of gravitational waves.

\subsection{Scalar perturbations in non-barotropic fluids}
As we mentioned in Section \ref{sflux}, if the symmetry  $\Phi^0\to \Phi^0+f(\Phi^a)$ is imposed on the general superfluid master function $U(b,X,Y)$, the operator $X$ drops and the resulting $U(b,Y)$ system is a perfect fluid. This type of perfect fluid is generically non-barotropic because the pressure and the energy density cannot be written one as a function of the other, see \eq{nbar}. This kind of action can be used to describe a generic perfect fluid, since $b$ can be identified with the entropy density and $Y$ with the chemical potential.\footnote{This thermodynamic interpretation
is not unique: $b$ can alternatively be associated to particle number
  density, and $Y$ to temperature,  see \cite{Ballesteros:2016kdx}. However, we stress that the results of this section are independent of this choice and can be derived purely algebraically, without resorting to any of the two possibilities.} Then, the entropy per particle $\sigma=b/U_Y$ is non-linearly conserved in
time \cite{Ballesteros:2016kdx}. The pressure perturbation can be written as
\be
\delta p = c_s^2 \, \delta \rho + \frac{\phi'}{a^4} \left(c_s^2-c_b^2 \right)\, \delta \sigma
\,, \qquad c_s^2 = \frac{p'}{\rho'} \,,\qquad  c_b^2 = \frac{\de p}{\de \rho} \Big |_b  \,,
\ee
\label{deltap}
where $\phi=\phi(t)$ describes the background evolution of the field $\Phi^0$, see \eq{cbasis}. In this expression and in all that follows, primes denote derivatives with respect conformal time $t$. The quantities $c_s^2$ and $c_b^2$ are, respectively, the standard adiabatic sound speed (i.e.\ the variation of the pressure with respect to the density at fixed $\sigma$) and the variation of the pressure with respect to the density at constant $b$. The latter of the two determines the evolution of $\phi$, which obeys (from the conservation of the EMT): 
\be
\phi'' + {\cal H} (3 \, c_b^2 -1) \phi' =0 \,. 
\label{phidd}
\ee

Let us now write the scalar part of the perturbed metric in the conformal Newtonian gauge:
\be
ds^2 =-a^2(t) \left(1-\psi\right) dt^2+a^2(t) \left(1+
  \tau\right) \delta_{ij} dx^i dx^j  \, .
\label{Newpert}
\ee
We will work in Fourier space, so that the perturbations $\psi$ and $\tau$ will be functions of conformal time and the modulus, $k$, of the comoving momentum $k^i$. Being $U(b,Y)$ a perfect fluid, there is no anisotropic stress and, in consequence, $\tau=\psi$. The evolution of the metric perturbation is then governed by the equation 
\be
\psi ''+3 \mathcal{H} \left(c_s^2+1\right) \psi'  +
 \left[c_s^2 \, k^2 +(3  \, c_s^2 +1) \mathcal{H}^2+2
   \mathcal{H}'\right] \psi  - \frac{\phi'  (c_s^2 -c_b^2)}{2 \,  a^2
   \, {M_{pl}}^2}
 \, \delta \sigma=0 \,.
\label{eqpsi}
\ee
We recall that the quantity $\Gamma\equiv\delta p - c_s^2 \, \delta \rho$ (that is sometimes called ``intrinsic entropy perturbation'') is gauge invariant and $\psi$ corresponds to one of the (gauge invariant) Bardeen potentials \cite{Bardeen:1980kt}. 
{The equation \eq{eqpsi} is a linear combination of the 0--0 and the $i$--$i$ Einstein equations.}
% For comparison, we also recall that for a $\Lambda$CDM universe the equation \eq{eqpsi}  reduces to 
%\begin{align} \label{lcdm}
%\psi ''+3 \mathcal{H} \psi'  +\left( \mathcal{H}^2+2\mathcal{H}'\right) \psi =0\,.
%\end{align}
 In the case at hand, {it} contains a generic time-dependent speed of sound  $c_s^2$, 
 %(which modifies the homogeneous part) %of \eq{eqpsi}
plus a time-dependent source term that is due to the non-barotropic character of $U(b,Y)$.

Indeed, whereas $\delta \sigma\equiv \delta\sigma(k)$ is constant in time, with its $k$-dependence fixed by initial conditions, the factor that multiplies it in equation \eq{eqpsi} has a non-trivial  time dependence determined by $c_s^2$ and $c_b^2$; which in turn depend on the function $U(b,Y)$.  Dynamical (and thermodynamic) stability of the perturbations requires $c_s^2>0$; see \cite{Ballesteros:2016kdx}. 

In  the special case of non-zero and constant $c_b^2$, from
(\ref{phidd}) one has that
\be \label{spc}
\phi' ={\phi_0'} \, a^{1-3 \, c_b^2}\,,
\ee
where {$\phi_0'$} is a constant {and we take $a=a_0=1$ today.} Then, the solution of \eq{eqpsi} during a matter domination period, where 
{$\mathcal{H}\propto 1/\sqrt{a}$,}
%$a\propto t^2 $, 
is given by
\begin{align}
\psi ={\tilde{\psi}}  -\frac{{\phi_0^\prime\,\delta \sigma}  }{3\,{M_{pl}^2}\,H_0^2\, \Omega_m^0 (6\,c_b^2-5) c_b^2}\, a^{-3 \, c_b^2},
\end{align}
where  {$\tilde\psi=\tilde\psi(k)$} is the constant solution of the homogeneous
equation {and $\Omega_m^0$ and $H_0=\mathcal{H}_0$ are, respectively, the relative matter density  and the Hubble parameter in cosmic time, both evaluated today.} We have neglected the decreasing solution of (\ref{eqpsi}) and {the effect of}
$c_s^2$, which is assumed to be positive but sufficiently small to be negligible for dark
matter. 

%
%When $c_b^2< -1/3$ the inhomogeneous (entropic) term of \eq{eqpsi} will increase as $a^{-(1+3c^2_b)}$ in a expanding Universe, affecting %significantly the growth of structure at late times, which is thus governed in this example by the equation 
%\begin{align}
%\psi ''+3 \mathcal{H} \psi'  + \left(\mathcal{H}^2+2
%   \mathcal{H}'\right) \psi  \propto  a^{-(1+3 \, c_b^2)}\,,
%\end{align}
%to be compared to the $\Lambda$CDM case \eq{lcdm}. 
%Modelling dark matter as a non-barotropic perfect fluid described by  $U(b,Y)$, 
The energy density perturbation can be written directly from the $0$-$0$ Einstein equation:
\be
\delta \rho = 
2 \, a^{-2} \, {M_{pl}^2} \left[\psi  \left(k^2+3 \,
    \mathcal{H}^2 \, \right)+3 \, \mathcal{H} \, \psi'\right]
\label{deltarho}
\ee
We get from it the time evolution of the {density contrast} $\delta = \delta \rho/\rho\,$:
\be
\label{dd}
\delta= \delta_{ad}-\frac{{\hat\sigma}}{3 (6\,c_b^2-5) c_b^2} \left({1-3\,  c_b^2}+\frac{{a} \, k^2}{3\,H_0^2 \, \Omega _m^0}\right)\, a^{-3\,  c_b^2}\,,
   \ee
{where 
\be 
\delta_{ad}=\tilde\psi
   \left({1}+\frac{{a} \, k^2}{{3}\,H_0^2 \, \Omega _m^0}\right)
\ee
is the non-entropic (or adiabatic) part of the fluctuation and we define the constant
\begin{align}
\hat\sigma\equiv \frac{\phi_0^\prime\,\delta \sigma  }{M_{pl}^2\,H_0^2\, \Omega_m^0 }\,. 
\end{align}}
%  \ee
%\be
%\label{dd}
%\delta= \tilde\psi
%   \left({1}+\frac{{a^2} \, k^2}{{3}\,H_0^2 \, \Omega _m^0}\right)-\frac{{\phi_0^\prime\,\delta \sigma}  }{3\,{M_{pl}^2}\,H_0^2\, \Omega_m^0 (6\,c_b^2-5) c_b^2} \left({1-3\,  c_b^2}+\frac{{a^2} \, k^2}{3\,H_0^2 \, \Omega _m^0}\right)\, a^{-3\,  c_b^2}\,.
%  \ee
%\be
%\label{dd}
%\delta= \psi _0
%   \left(\frac{2 \, a \, k^2}{H_0^2 \, \Omega _m}+6\right)+ \frac{
%     a^{-3\,  c_b^2} \left[6 \, H_0^2 \left(3
%   c_b^2-1\right) \Omega _m-2 \, a \, k^2\right]}{3 \, c_b^2 \, 
%  \,  H_0^4 \left(6 \, c_b^2-5\right) \Omega _m^2} \delta \hat \sigma
% 2 \;\psi_0+\frac{2(1-3\,c_b^2)\;\phi_0\;\delta\sigma}{3(5-6\,c_b^2)\;{M_{pl}^2}\,\mathcal{H}_0^2}\;a^{-3\,c_b^2}+\frac{2\;k^2\;a}{3\;\mathcal{H}_0^2}\left(
% \psi_0+\frac{\phi_0\;\delta\sigma}{3\,(5-6 \,c_b^2)\;{M_{pl}^2}\,\mathcal{H}_0^2}\;a^{-3\,c_b^2}
% \right)
%\,;
%\ee
% where we have defined $\hat \sigma = \phi_0 \, M_{pl}^{-2}\,  \sigma$. 
With respect to the standard cold dark matter (for which $\delta\sigma=0$),  the entropic term modifies the time evolution of the density contrast.
%, $g= \delta/a$ in a $k$-dependent way.
It is interesting to note that as soon as {$c_b^2<0$}, there is an entropic mode that grows with respect to the  adiabatic one, both for super- and sub-horizon scales.\footnote{Notice that in a more general case, where $c_s^2$ is not negligible, there is an interplay between $c_s^2$ and $c_b^2$ that makes the growth condition more complicated that just $c_b^2<1$.}
%Assuming that the background pressure of the dark matter vanishes (which can be ensured with an adequate choice of the function $U$), and neglecting for the subdominant contributions from baryons and photons during matter domination, the equation for the time evolution of the {density contrast} $\delta = \delta \rho/\rho$ for $k\gg\mathcal{H}$ can be written, using (\ref{deltarho}) and (\ref
%{eqpsi}), as 
%
%\be
%\delta ''+\mathcal{H} \, \delta '{-\frac{3}{2}  \,  \mathcal{H}^2 \, 
  % \delta}- \frac{c_b^2\,k^2 }{6} \frac{a\, \phi '}{M_{pl}^2\,\mathcal{H}^2}\, \delta \sigma 
%=0 \,.
%\ee
%With respect to the standard cold dark matter (for which $\delta\sigma=0$),  the entropic term modifies the time evolution of the grow function, $g= \delta/a$ in a $k$-dependent way.
As an explicit example we can take the function
$U(b,Y)$ corresponding to a perfect non relativistic
gas~\cite{Ballesteros:2016kdx}:
\be
U(b, \, Y) =  b Y \left \{ 1+ \text{Log} \left[\frac{g_s}{b}\left( \frac{m
    Y}{2 \pi} \right)^{3/2} \right] \right \} - b\, m \,,
\ee
where $g_s$ is the number of spin states and $m$ is the atomic mass of
the gas. One gets
%\be
$c_b^2 = {{2}/{3}} $
%\ee
%
so that {$\phi'\propto a^{-1}$, see \eq{spc}, and
\be
\label{dds}
\delta= \delta_{ad}+\frac{\hat \sigma  }{2} \left(-1+\frac{a \, k^2}{3\,H_0^2 \, \Omega _m^0}\right)\, a^{-2}\,.
  \ee}
%\be
%\delta=\psi _0
%   \left(\frac{2 \, a \, k^2}{H_0^2 \, \Omega _m}+6\right)
%-\frac{ \left(6 \, H_0^2 \, \Omega _m-2 \,  k^2 \, a\right)}{2 \, 
%   H_0^4 \, \Omega _m^2 \, a^2} \delta \hat \sigma 
% 2 \;\psi_0-\frac{2\;\phi_0\;\delta\sigma}{3\;{M_{pl}^2}\,\mathcal{H}_0^2\;a^2}+
% \frac{2\;k^2\;a}{3\,\mathcal{H}_0^2}\left(
% \psi_0+\frac{\phi_0\;\delta\sigma}{3 \;{M_{pl}^2}\,\mathcal{H}_0^2\;a^2}
% \right)
%\, ;
%\ee
%
{Therefore, in this example}  the entropic corrections {decrease} in time with respect the adiabatic perturbation (since $c_b^2>0$).

A more dramatic example {of} the influence of $\delta \sigma$ is
obtained, for instance, by taking %with a simple master function such as
\be \label{dedm}
U(b, \, Y) =- \lambda \;b+V(b\,Y) \, 
\ee
%
%with $V$ a generic function, that merges in a single fluid the dark matter and dark energy features.
In this case, $c_b^2=-1$ and {$\phi'\propto a^{4}$, see \eq{spc}, so that}
 $b\,Y={\phi_0^\prime}$ is constant in time, and 
\be
c_s^2=0, \, \qquad
w=p/\rho= {\frac{V(\phi_0^\prime)-\phi_0^\prime\;\frac{dV}{d(bY)}\big{|}_{\phi_0^\prime}}{{\lambda}/{a^3}-V(\phi_0^\prime)+\phi_0^\prime\;\frac{dV}{d(bY)}\big{|}_{\phi_0^\prime}}}
\, .
\ee
%-\frac{\phi_0^2 \,a^3\,V''(\phi_0)}{\lambda+\phi_0^2 \,a^3\,V''(\phi_0)}$
In the regime {in which $\lambda/a^3$ dominates the denominator, $w \rightarrow 0$ (as cold dark matter) in a first approximation. In the opposite limit, $w\rightarrow -1$ and the fluid describes a phase of dark energy domination.}
% \be
% w=\frac{V(\phi_0)-\phi_0\;V'(\phi_0)}{{\lambda}/{a^3}-V(\phi_0)+\phi_0\;V'(\phi_0)}\,,
% \ee
%corresponds to dark matter domination with the matter constrast
% described by \eq{dd}. When $\lambda\ll
% \,a^3\,(V(\phi_0)-\phi_0\;V'(\phi_0)$, the fluid has an equation of
% state $w\sim -1$ and describes dark energy domination
During matter domination, when $w\sim 0$, the energy density contrast is given by {
\be
\delta= \delta_{ad}-\frac{\hat\sigma }{33} \left(4+\frac{a\, k^2}{3\,H_0^2 \, \Omega _m^0}\right)\, a^{3}\,.
  \ee}
In this case, {the} entropic {perturbation} grows faster than the adiabatic one.
%\footnote{Notice that $a^3$ grows during matter domination by a factor $10^{12}$! {???}} 
Although {in this example,} in order to keep under control the validity of linear perturbation theory, very strong constraints have to be imposed on the initial conditions {(specifically on $\hat\sigma $),} it illustrates the relevance of possible interactions between dark matter and dark energy in the context of the effective theory of self-gravitating media. 

In general, by splitting $U(b,Y)$ {as} $U_{DM}(b,Y) + V(b \, Y)$ one can
systematically study the coupling of dark matter and dark energy,
being the corresponding EMT tensors not separately conserved.

{To conclude this subsection} it is also worth hinting how the evolution of perturbations  is affected for a superfluid and a solid. In the first of these two cases, described by a function $U(b,X,Y)$, the function $\delta \sigma$ is not constant in time
and the conservation equation $\delta\sigma'=0$ will be replaced by a dynamical one. On the other hand, if a solid is considered, an anisotropic stress is present  and the two Bardeen potentials are not equal, so that
\be \label{diffB}
 \tau- \psi = 4 M_2^2 \, {\pi_L} \, ,
\ee
where {$\pi_L$} is the longitudinal mode of the spatial St\"uckelberg perturbations: {$\Phi^i= x^i +
  \de_i \pi_L$, }see also \eq{anis}. We plan to do a complete  analysis of both effects in a forthcoming paper.
  
{\subsection{Gravitational waves}
The dynamics of gravitational waves is governed by \eq{GW}. Without loss of generality, we can choose $N(t)=a(t)$, so that in Fourier space we have the action
\begin{align}
\label{GW2}
\frac{M_{pl}^2}{2}\int dt\,d^3k\, a^2 \left[\chi_{ij}'
\chi_{ij}'+
\left( 2 \, M_{2}^2  + k^2 \right) \chi_{ij}  \,
\chi_{ij}\right ]  \,.
\end{align}
The resulting propagation equation is
\begin{align}
\chi_{ij}''+2\mathcal{H}\chi_{ij}'+(k^2-M_2^2)\chi_{ij}=0\,.
\end{align}
The difference with respect to the standard expression is the presence of the time-dependent effective mass squared, $M_2^2$, which is non-vanishing for solids and supersolids. As we had anticipated, this mass also controls the difference of the Bardeen potentials, see \eq{diffB}. We point out that, at leading order in derivatives (LO), $M_2^2$ induces a modification of the speed of propagation (the dispersion relation) of gravitational waves, which becomes scale-dependent. Interestingly, it was shown in \cite{Ballesteros:2014sxa} that at next-to-leading order (NLO), the propagation speed changes in a scale-independent (but time-dependent) way in the case of $V$Diff invariant media made of just the three spatial St\"uckelberg fields. The same kind of modification is expected to occur at NLO for a generic medium, leading to a generic propagation equation of the form:
\begin{align}
\chi_{ij}''+2\mathcal{Q}_1\mathcal{H}\chi_{ij}'+(\mathcal{Q}_2\,k^2-\tilde M_2^2)\chi_{ij}=0\,,
\end{align}
where $\tilde M_2^2$, $\mathcal{Q}_1$ and $\mathcal{Q}_2$ are time-dependent functions and we expect $M_2^2/\tilde M_2^2\sim 1+\mathcal{O}(\mathcal{H}/\Lambda)$ and also $\mathcal{Q}_1,\mathcal{Q}_2\sim 1+\mathcal{O}(\mathcal{H}/\Lambda)$, where $\Lambda$ denotes the cut-off of the effective theory. 

}

\section{Conclusions and outlook} \label{conc}
The main result of this work is that a vast class of modified gravity models, including massive gravity (MG), can be interpreted as self-gravitating continuous media. The low-energy dynamics of these systems is described by the EFT of four scalar fields $\Phi^A$ that respect shift symmetries. Due to these symmetries, the fields are derivatively coupled between them and are minimally coupled to gravity at leading order in derivatives. At this order, the action is a functional of ten independent scalars encoded in the induced  four-dimensional  (inverse) metric $C^{AB}=g^{\mu\nu}\partial_\mu\Phi^A\partial_\nu\Phi^B$. 

The four scalar fields can be interpreted as the (comoving) coordinates of the medium. They can be conveniently split into three spatial coordinates $\Phi^a$, $a=1,2,3$, and a temporal coordinate $\Phi^0$. Using diffeomorphism (diff) invariance, the EFT can be examined in the unitary gauge, where the scalar fields are ``frozen'' to be coincident with the spacetime coordinates. With this choice of spacetime coordinates, the induced metric is $C^{AB}=g^{AB}$. This sends all the dynamics of the medium into the gravitational metric and allows to make direct contact with the traditional framework of MG. The inverse path (going from MG to continuous media) can be travelled by using the well-known St\"uckelberg ``trick'', which allows to write in a diff invariant way any theory in which diffs appear to be broken. Indeed, the four scalar fields $\Phi^A$ of our EFT can also be seen as the four St\"uckelberg fields of a diff invariant formulation of MG. We have discussed in Section \ref{massg} how to move from one picture to the other.

We have provided a comprehensive and systematic classification of massive gravity and modified gravity models in terms of symmetries and low-energy degrees of freedom, establishing a direct connection to the theory of (self-gravitating) continuous media and the pull-back formalism. In doing so, we have determined also which are the propagating degrees of freedom for each case (as a function of the symmetries) and obtained the Lagrangians that are necessary for the study of (linear) cosmological perturbations at large scales, highlighting which operators affect the different kinds of observables.

The mechanical and thermodynamic properties of these continuous media (or modified/massive gravity) models depend crucially on extra symmetries that can be imposed on the scalar sector and may restrict drastically  the number of allowed operators. The minimal assumption we have used in this work is that the action is invariant under (spatial) SO(3) rotations of the fields $\Phi^a$, $a=1,2,3$. This reduces the number of allowed operators from ten to nine at leading order in derivatives; see  Table \ref{tab:operators}. Imposing further assumptions, such as symmetries that relate the spatial and temporal St\"uckelberg sectors --see Table \ref{tab:summa}--, the number of independent operators can be decreased even down to a single one. For instance, volume preserving internal spatial diffs selects just three operators and the resulting action describes superfluids. 

Another way to simplify the action is to make it depend (by assumption) only on either the spatial or temporal St\"uckelberg fields; see Table \ref{3dimmedia}. The models that are allowed in those cases can describe self-gravitating solids, perfect fluids and superfluids. In the general SO(3) spatially symmetric case, which contains the four scalars $\Phi^A$, the medium shares some of its properties with both superfluids and solids, and therefore is called a supersolid. Therefore, a general MG model is interpreted a supersolid propagating in spacetime. 

Working in the unitary gauge and expanding a generic example of our EFT around Minkowski spacetime, it is straightforward to see that the resulting action for the metric fluctuations will not respect Lorentz symmetry. This is simply a natural consequence of the presence of the medium and the fact that we have assumed the operator content to be dictated by a broad symmetry: internal spatial SO(3) rotations. Nevertheless, it is possible to choose a combination of operators such that the resulting action for metric perturbations is Lorentz invariant, if one wishes to do so \cite{nonder-us}.

The correspondence between self-gravitating media and massive/modified gravity is intriguing and deserves further study. In the context of cosmology, it can be relevant for a deeper understanding of the properties of dark matter and dark energy. Moreover, the systematic EFT framework presented in this work can aid to devise a program for the interpretation of the data coming from future probes. From this perspective, the key idea is that the symmetry properties of a (coarse-grained) cosmological medium could be imprinted in the the large scale structure of the Universe, including the cosmic microwave background. 

We plan to analyze the cosmology of these models in a future publication, by studying in further detail the signatures that would allow to distinguish these models from the standard $\Lambda$CDM model \cite{future-cosm}. In the present work, we have initiated  this {program (in Sections \ref{cosmox} and \ref{sigs}) by determining the number of propagating degrees of freedom in FLRW for various representative examples of continuous media (or models of massive/modified gravity), as summarized in Table \ref{T3}, and by studying analytically some representative examples. In particular, in Section \ref{sigs} we have considered the modification to the growth function of matter in non-barotropic perfect fluids and the modifications to the propagation of gravitational waves.} Other  avenues, that are also worthwhile exploring, can be opened if different internal symmetries --such as supersymmetry \cite{Hoyos:2012dh} or Galileon symmetry \cite{Nicolis:2015sra}-- are assumed instead of spatial SO(3). 

\section*{Acknowledgements}
We thank B.\ Bellazzini and A.\ Ramos for discussions. We thank D.\ Blas {and S.\ Sibiryakov} for useful comments on the draft. D.C.\ and L.P.\ would also like to thank F.\ Nesti for enlightening discussions during previous works.
The work of G.B.\ is funded by the European Union's Horizon 2020 research and innovation programme under the Marie Sk{\l}odowska-Curie grant agreement number 656794. G.B.\ thanks the DESY theory group for hospitality while part of this work was developed and the German Science Foundation (DFG)  for funding through the  Collaborative Research Center (SFB) 676 ``Particles, Strings and the  Early Universe''. G.B.\ thanks as well the CERN Theoretical Physics Department for hospitality while this work was completed. G.B. and L.P. would like to thank Nordita (Nordic Institute for Theoretical Physics) for hosting them during a fruitful workshop. L.P.\ thanks  the Institute de Physique Th\'eorique IPhT CEA-Saclay for hospitality during the preparation of this work.

\begin{appendix}

\section{Conserved currents and charges} \label{conse}

Each of the systems we consider in this paper is characterized by some conserved currents and their associated charges. These currents and charges have special significance for the thermodynamic and dynamical interpretation of the different media. Generically, the currents can be broadly classified in two different types.
First, we have those currents which, by virtue of Noether's theorem, are due to the symmetries that define the Lagrangians of the different models.  These currents are grouped in sets of infinite dimension because the defining symmetries are (infinite) subgroups of the internal diffeomorphisms $\Phi^A\to\Psi^A(\Phi^B)$. Then, we also have currents that are  conserved irrespectively of the equations of motion. An example  of the latter is  
\begin{align} \label{fec}
J^\mu=-b\, u^\mu\,,
\end{align} 
where $u^\mu$ was defined in \eq{veloc}. This current is covariantly conserved off-shell and exists for all the models we consider (regardless of their symmetries) that involve the spatial St\"uckelberg fields $\Phi^a$. It has often been interpreted as the entropy current for fluid actions of the form $S=\int d^4x\, U(b)$, see e.g.\ \cite{Dubovsky:2011sj,Ballesteros:2012kv}, although this is not the only possible interpretation, see e.g. \cite{Schutz:1970my}. 

Independently of whether a current is conserved due to a symmetry of the action or for another reason, an associated charge that is conserved in time can be defined. In general, given a current $J^\mu$ that satisfies 
\begin{align}
\nabla_\mu J^\mu=0\,,
\end{align} 
the corresponding time-conserved charge is 
\begin{align}
Q=\int d^3x\, \sqrt{-g }\, J^0=\int d^3\Phi\frac{J^0}{b\, u^0}\,,
\end{align} 
which can be proven using the divergence theorem and assuming that the fields fall of quickly enough at infinity. 
 For instance,  in the particular case of the current \eq{fec}, the charge is
\begin{align}
\mathcal{V}_3=-\int d^3\Phi\propto \epsilon_{ijk}\,\epsilon^{0\alpha\beta\gamma}\int d^3x\, \partial_\alpha\Phi^i \partial_\beta\Phi^j \partial_\gamma\Phi^k
\end{align}
and, clearly, $\mathcal{\dot V}_3=0$ due to the antisymmetric character of $\epsilon^{\mu\alpha\beta\gamma}$. The physical interpretation of this result is that the flux lines of the medium are neither created nor destroyed, which is a topological statment. In other words, the conservation of $\mathcal{V}_3$ means that these continuous media do not have flux sources nor sinks. 

\subsection{General volume currents}

Let us first consider conserved currents that are independent of the internal symmetries. We just encountered one example of them in \eq{fec}. This type of current can be generalized in a simple way with a permutation of the fields $\Phi^A$. Concretely, the four currents 
\begin{align} \label{4currents}
J^\mu_D=\frac{\epsilon^{\mu\alpha\beta\gamma}}{6\sqrt{-g}}\,\epsilon_{ABCD}\,\partial_\alpha\Phi^A\partial_\beta\Phi^B\partial_\gamma\Phi^C\,,\quad D=0,1,2,3
\end{align}
are covariantly conserved, and $J^\mu_0$ is precisely  \eq{fec}. In fact, any four-vector of the form
\begin{align}
\mathcal{J^\mu}_{D}[h(\Phi^A)]=h(\Phi^A)\, J^\mu_D\,,
\end{align}
where $h$ is an arbitrary function of the  fields $\Phi^A$, satisfies
\begin{align} \label{ide}
\nabla_\mu \mathcal{J^\mu}_{D}[h(\Phi^A)]= 4\frac{\partial h(\Phi^A)}{\partial\Phi^D}\det\left(\frac{\partial\Phi^B}{\partial x^\nu}\right).
\end{align}
This allows to construct further conserved currents by simply multiplying any of the currents $J_D^\mu$ with a function $h$ that does not depend on the field $\Phi^D$. The relation \eq{ide} can be easily proven using the identity $\sqrt{-g}\,\nabla_\mu U^\mu=\partial_\mu\left(\sqrt{-g}\,U^\mu\right)$, valid for any vector $U^\mu$, and the antisymmetry of $J^\mu_D$ under the permutation of two fields $\Phi^A$.
  
\subsection{Noether currents} \label{noth}

Let us now consider the currents coming from the different symmetries that we have considered in the paper. First of all, we have $SO(3)_s$ and the internal shifts \eq{ssyim}. The first one of these two symmetries leads to the conservation of angular momentum in the internal spatial manifold. The conservation of the currents associated to the shift symmetries is nothing but the equations of motion of the St\"uckelber fields, see \eq{gseqm}.  We will now enumerate the conserved currents due to the additional symmetries of Section \ref{wecf}.

\subsubsection{Four-dimensional media} 

\begin{itemize}

\item If the action is invariant under $V_s\text{Diff}$, the volume-preserving spatial diffeomorphisms in the space of St\"uckelberg fields, the operators $b$, $Y$ and $X$ are selected and the resulting LO medium is a superfluid. The currents associated to this symmetry are known to be related to vorticity conservation, see for instance~\cite{Dubovsky:2005xd,Ballesteros:2012kv}. These currents are of the form:
\begin{align}
J^\mu_{(\varepsilon)} = -b\, U_b(\boldsymbol{B}^{-1})^{cd}\partial^\mu\Phi^d\varepsilon^c(\Phi^j)\,,
\end{align}  
where $\varepsilon^a(\Phi^j)$ satisfies $\partial \varepsilon^a/\partial \Phi^a=0$\,. A basis of vorticity charges (in which any other vorticity charge can be expressed) is constructed choosing $\varepsilon^a =\epsilon_{abc}\,\alpha_b\,\partial \delta^3(\Phi-\tilde\Phi)/\partial \tilde\Phi^c$, with constant $\alpha_b$, see e.g.\ \cite{Ballesteros:2012kv}.

\item The covariantly conserved Noether currents associated to the symmetry $\Phi^a\rightarrow \Phi^a+f^a(\Phi^0)$, which selects the operators $X$ and $w_n$, are:
\begin{align}
J^{\mu}_f=\sum_{a=1}^3f^a\,(\Phi^0)\sum_{n=1}^3 n\, U_{w_n}\left[\boldsymbol{W}^{n-1}\cdot j\right]^{\mu}_a\,,
\end{align}
where we define
\begin{align}
j^\mu_a=\left(\partial^\mu\Phi^a-\mathcal{C}^a X^{-1}\partial^\mu\Phi^0\right)\,.
\end{align}
Each set of three functions $\{f^1,f^2,f^3\}$ of the temporal St\"uckelberg field defines a different current. By using the unitary gauge expressions \eq{asfll} we interpret
$\Phi^a\rightarrow \Phi^a+f^a(\Phi^0)$ as a transformation that does not change the slicing of spacetime but modifies the threading with a time-dependent shift: $N\rightarrow N$\,, $N^i\rightarrow  N^i+N^{-1}{\partial f^i}/{\partial\Phi^0}$. Notice the spatial part of the metric, $\gamma_{ij}$, is invariant under $\Phi^a\rightarrow \Phi^a+f^a(\Phi^0)$.

\item In the case of the symmetry $\Phi^0\rightarrow \Phi^0+f(\Phi^a)$, the symmetric LO scalar operators are $Y$ and $\tau_n,$ and the covariantly conserved currents are:
\begin{align} \label{c0}
J^\mu_f=f(\Phi^a)\,Y\,U_Y\,u^\mu\,.
\end{align}
It seems natural to interpret this set of currents --recall that there is a current for each choice of the function $f$-- as transporting charges of the fluid in the direction of the four-velocity $u^\mu$. Notice that these currents are parallel to the current \eq{fec}. Therefore, if \eq{fec} is interpreted as the entropy current, the currents \eq{c0} carry charges that flow with the entropy. 

\item If the symmetry $\Phi^0\rightarrow \Phi^0+f(\Phi^0)$ is imposed, the LO action depends on the operators $\tau_n$ and $\mathcal{O}_{\alpha\beta n}$. The operators $\tau_n$ do not contribute to the conserved currents, since they do not contain $\Phi^0$. The currents in this case are
\begin{align}
J^\mu_f=\frac{f(\Phi^0)}{Y^2}\sum_{\alpha,\,\beta,\,n}U_{\mathcal{O}_{\alpha\beta n}}
\end{align}
where the sums extend over all the $\mathcal{O}_{\alpha\beta n}$ operators and
\begin{align}
j^\mu_{\alpha\, \beta\, \gamma}=\alpha\,\mathcal{O}_{(\alpha-1)\beta n}\,\partial^\mu\phi^0-(\alpha+\beta)\,Y\,\mathcal{O}_{\alpha\beta n} u^\mu+\beta\,\mathcal{O}_{\alpha(\beta-1) n}\left(\boldsymbol{B}^n\right)^{ab}C^{0a}\partial^\mu\Phi^b\,.
\end{align}
As we discussed in Section \eq{wecf}, the symmetry $\Phi^0\rightarrow \Phi^0+f(\Phi^0)$ allows at LO an infinite number of different scalars $\mathcal{O}_{\alpha\beta n}$. This is because the exponents $\alpha$ and $\beta$ appearing in the definition of $\mathcal{O}_{\alpha\beta n}$ are arbitrary real numbers (whereas $n$ can take the values 1, 2 or 3). In consequence the summatory $\sum_{\alpha,\,\beta,\,n}$ has to be understood to extend over all the possible values of these parameters, unless some restriction is enforced on them. 

\end{itemize}

\subsubsection{Media with reduced internal dimensionality}

\begin{itemize}

\item We can also consider the case in which the Lagrangian depends only on the temporal St\"uckelberg field, although, as we explained is Section \eq{UdX}, we do not know of any symmetry that forbids the spatial St\"uckelberg fields at LO (and allows them at higher orders). In this case the master function $U$ depends exclusively on $X$ at LO and the relevant symmetry is the invariance under $\Phi^0\to\Phi^0+c^0$, with $\partial_\mu c^0=0$. This simply tells us that the current
\begin{align} \label{xshift}
\mathcal{X}^\mu =\sqrt{-X}\,U_X \vel^\mu
\end{align}
is covariantly conserved. The conservation of $\mathcal{X}^\mu$ is precisely the equation of motion for $\Phi^0$, given the master function $U(X)$. 

\item Analogously, we can consider the case in which only the spatial St\"uckelberg fields are present. Since we are assuming always an $SO(3)_s$ symmetry this corresponds to the standard solid  $U(\tau_n)$. Notice, again, that we know of no symmetry that prevents the appearance of $\Phi^0$ at LO but reintroduces this field at higher orders. Aside from the conservation of internal angular momentum, due to $SO(3)_s$, there are also the currents associated to the internal translational invariance $\Phi^a\to\Phi^a+c^a$, with $\partial_\mu c^a=0$. This symmetry generates in this case the set of covariantly conserved currents
\begin{align} \label{tshift}
t^\mu=c^a\sum^3_{n=1}n\,U_{\tau_n}\left(\boldsymbol{B}^{n-1}\right)^{ab}\partial^\mu\Phi^b\,,
\end{align}
a base of which is obtained by setting, alternatively, $c^a$ to 0 or 1 for $a=1$, 2 or 3. 

The currents associated to $SO(3)_s$ invariance have a similar structure. In fact, they can be formally obtained from the previous ones simply replacing $c^a$ by $r^{ab} \Phi^b$, where $r^{ab}$ belongs to the Lie algebra of $SO(3)_s$, and so is antisymmetric. These are just the generators of the angular momentum in the internal spatial manifold. Therefore, the currents
\begin{align}
J_r^\mu=r^{ab}\Phi^b\sum^3_{n=1}n\,U_{\tau_n}\left(\boldsymbol{B}^{n-1}\right)^{ab}\partial^\mu\Phi^b\,,
\end{align}
are also conserved in this case. A basis for this currents can be obtained choosing the matrices $r^{ab}$ as the standard ${L_x,L_y, L_z}$ generators. The conservation of the currents of this basis are the equations of motion for the spatial St\"uckelberg fields.

\end{itemize}

\section{A shortcut to count degrees of freedom} \label{letus}

We can determine the number of DoF by studying directly to the equations of motion of the St\"uckelberg fields (\ref{gseqm}). From the action \eq{saction} we know that there are at most six DoF, that is: two from the metric plus those coming from the scalar fields ($\leq 4$). The equation (\ref{gseqm}) can be rewritten in the following form (using the notation $U_{AB} =\partial U/\partial C^{AB}$):
\be \label{geneq}
{\cal L}_A=\left[U_{AB}\;g^{\mu\nu}+\,(U_{AD,CB}+U_{AD,BC})\,\nabla^{\mu}\,\Phi^C\,\nabla^{\nu}\,\Phi^D\right]\,
\nabla_{\mu}\nabla_{\nu}\,\Phi^B\equiv{\mathcal K}^{\mu\nu}_{AB}\;\nabla_{\mu}\nabla_{\nu}\,\Phi^B =0,
\ee
where ${\mathcal K}^{\mu\nu}_{AB}$ contains only first derivatives of the
fields. From this expression we see that the second order time derivatives of $\Phi^A$ are always proportional to
the matrix kernel ${\mathcal K}^{00}_{AB}$, thus the number of St\"uckelberg propagating DoF is 
\be\label{dof}
\Delta = {\rm Rank}[ {\mathcal K}^{00}_{AB}]\,.
\ee
To simplify the analysis it is convenient to use the phonons \eq{phonons} and study the system perturbatively. The coefficient of second order time
derivatives of the perturbations are determined by  $\bar {\mathcal
  K}^{00}_{AB}\equiv{\mathcal K}^{00}_{AB} {}_{| \Phi^A=x^A}$. From \eq{geneq} we have that
$(\bar{\mathcal K}^{00}_{AB} +{\cal O}(\partial\pi))\ddot\pi^B=0$ where
\be
\bar{\mathcal K}^{00}_{AB}=\bar{L}_{AB}\;g^{00}+\,(\bar{L}_{AD,CB}+\bar{L}_{AD,BC})\,g^{0C}\,g^{0D}\,,
\ee
where the overbars mean that the corresponding quantities are evaluated on $\Phi^A = x^A$. It is easy to see that $\bar {\mathcal K}^{00}_{AB}$ gives
the number of DoF also at the non perturbative level. Indeed, by going to the unitary gauge we have that ${\rm Rank}[ {\mathcal K}^{00}_{AB}]=\Delta$ and so the counting  of DoF is background-independent. For $SO(3)_s$ invariant Lagrangians one finds the same results obtained applying the
Hamiltonian formalism \cite{canonical-us,general-us,nonder-us}, that is:
\begin{itemize}
\item[$-$] for $U=L(g^{AB})$ we get $ 4\;{\rm scalar}\; \text{DoF}$ 
\item[$-$] for $U=L(g^{ab})+\sqrt{-g^{00}}\;\tilde{ L}(\,\gamma^{ab}) $ we
  get $ 3\;{\rm scalar}\;\text{DoF}$  
\item[$-$] for $L_1=L(g^{00},\,\gamma^{ab} )$ we get $ 1\;{\rm scalar}\;\text{DoF}$  
\item[$-$] for $L_0=\sqrt{-g^{00}}\;\tilde{ L}(\,\gamma^{ab})$ we get $
0 \;{\rm scalar}\;\text{DoF}$  
\end{itemize}
The Hamiltonian analysis applied to field theories allows for
a  non-integer number  of DoF (see \cite{canonical-us,nonder-us}) while the  prescription of (\ref{dof}) seems to evade such a problems.
For example the naive Hamiltonian counting of DoF for a master function of the form $U=\sum_a C^{0a}C^{0a}/C^{00}$ gives 3+1/2 DoF in the Hamiltonian formalism and 3 DoF when computed following (\ref{dof}). This mismatch is being studied.

%%%%%%%%%%%%%%%%

\end{appendix}

\end{document}